\documentclass[12pt,a4paper]{article}

\usepackage[english]{babel}
\usepackage[T2A]{fontenc}
\usepackage[cp1251]{inputenc}
\usepackage{amsmath}
\usepackage{graphicx}
\usepackage{amssymb}
\usepackage{color}
\usepackage{amsfonts}
\usepackage{wrapfig}
\usepackage{caption}

\usepackage[normalem]{ulem}

\usepackage[hyphens,spaces,obeyspaces]{url}
\usepackage[colorlinks,breaklinks,allcolors=blue]{hyperref}
\usepackage{doi}

\usepackage{physics}

\begin{document}
	
	\binoppenalty=10000
	\relpenalty=10000
	
\begin{center}
	\textbf{\Large{Thermodynamic response functions of the Curie-Weiss cell fluid model. I. Supercritical region}}
\end{center}

\vspace{0.3cm}

\begin{center}
{Oksana~A.~Dobush\footnote{e-mail:  dobush@icmp.lviv.ua}, Mykhailo~P.~Kozlovskii,  Roman~V.~Romanik, Ihor~V.~Pylyuk, Mykola~A.~Shpot} 
\end{center}

\begin{center}
	\textit{Yukhnovskii Institute for Condensed Matter Physics \\ of the National Academy of Sciences of
	Ukraine, 79011 Lviv, Ukraine}
\end{center}

	 \vspace{0.1cm}

	\begin{abstract}
\small In this work, we present an analytical study of the thermodynamic response functions of a multiple-occupancy cell fluid model with competing Curie-Weiss attraction and local repulsion. In the grand canonical ensemble, explicit expressions are derived for the three independent response functions: the isothermal compressibility, the thermal pressure coefficient, and the isochoric heat capacity. The thermal expansion coefficient and the isobaric heat capacity are subsequently obtained from exact thermodynamic identities. The analytical formulas are used to investigate the behavior of the response functions in the supercritical region. The isothermal compressibility, the thermal expansion coefficient, and the isobaric heat capacity exhibit critical divergences, whereas the thermal pressure coefficient and the isochoric heat capacity remain finite at the critical points despite developing pronounced extrema. The thermal pressure coefficient displays a non-monotonic oscillatory dependence on density and becomes negative over a limited range of temperatures and densities, leading to negative values of the thermal expansion coefficient. The obtained analytical expressions provide a consistent theoretical framework for investigating the thermodynamic properties of the model and establish a basis for future analysis of the subcritical regime and multiphase equilibrium. 

\vspace{0.1cm}

    \end{abstract}

	\normalsize 
\section{Introduction}

Thermodynamic response functions are fundamental for analyzing and interpreting the behavior of physical systems. They quantify how a system reacts to variations in its state parameters, offering deep insights into its thermodynamic characteristics. For fluids, the most commonly investigated response functions include the isothermal compressibility, the isobaric thermal expansion coefficient, the isochoric thermal pressure coefficient, and the isochoric and isobaric heat capacities. These quantities serve as indispensable tools for probing, predicting, and controlling the thermodynamic behavior of matter. They find extensive applications across scientific and engineering fields, where they aid in the modeling, optimization, and theoretical understanding of diverse processes~\cite{KDEetal21, SI23}. Consequently, considerable research attention is devoted to studying the thermodynamic properties of simple fluids and model systems, both theoretically and through simulations and experiments, in the subcritical and supercritical regimes~\cite{Johnston14,YS13,VRMBOA11,Liu23}.

Despite their central role in equilibrium thermodynamics, analytical studies of thermodynamic response functions remain rather limited for exactly solvable models of fluids. In many realistic systems these quantities are obtained by numerical differentiation of an equation of state~\cite{BDL22,Gholami25arxiv,Deiters25,DB21,KW20}, which may obscure their analytical structure and complicate the investigation of critical behavior. Exact expressions are therefore particularly valuable, as they provide direct insight into the interplay between microscopic interactions and macroscopic thermodynamic properties while serving as reliable benchmarks for approximate theoretical and numerical approaches~\cite{HD26,ZYHP22}.

Being a multiple-occupancy system, the cell fluid model (CFM) with Curie–-Weiss interaction provides an attractive framework for such an investigation. Owing to the competition between a global mean-field attraction and a local intracell repulsion, the model remains analytically tractable while exhibiting a remarkably rich phase behavior. Previous studies have demonstrated the existence of an infinite hierarchy of first-order phase transitions, with each coexistence curve terminating at a corresponding critical point~\cite{KKD18,KKD20,KD22}, as well as the emergence of triple point upon suitable modification of the interaction parameters~\cite{KDPRS26}. These features make the model particularly well suited for studying how thermodynamic response functions reflect complex phase behavior.

Multiple-occupancy lattice and cell models constitute an important extension of conventional single-occupancy lattice-gas models. {Multiple-occupancy lattice models are closely related to systems with soft or penetrable effective interactions, for which the energy cost of particle overlap remains finite}, for example the Gaussian core model~\cite{Stillinger76}, generalized exponential model of index 4 (GEM-4)~\cite{MGKNL06}, the penetrable spheres model~\cite{LWL98}. Such models are relevant to a broad class of soft-matter systems~\cite{Likos01}, where effective interactions may differ substantially from those characteristic of fluids with hard core exclusion. From the statistical-mechanical perspective, multiple-occupancy models are also of particular interest because they can exhibit complex collective behavior while retaining a comparatively simple mathematical structure.

The primary objective of the present work is to establish a complete analytical description of the thermodynamic response functions of the CFM in the supercritical region. We derive exact expressions for the three independent response functions: the isothermal compressibility, the thermal pressure coefficient, and the isochoric heat capacity, from which the remaining response functions follow through standard thermodynamic identities. Besides providing efficient formulas for numerical evaluation, these results clarify the relationship between the microscopic statistical description of the model and its macroscopic thermodynamic behavior.

The present paper constitutes the first part of a systematic investigation of the thermodynamics of the CFM with Curie–Weiss interaction. Here we focus on the supercritical region, where the analytical structure of the response functions can be established most transparently. The derived expressions provide the theoretical foundation for the future companion study devoted to the subcritical region, where phase coexistence, metastability, and the infinite sequence of first-order transitions will be analyzed.

The model describes an open system of interacting point particles confined in a three-dimensional volume $V$, which is partitioned into $N_v$ mutually disjoint, congruent cubic cells $\Delta_l$, $l\in\{1,...,N_v\}$, each of a volume $v$. 
The pairwise Curie-Weiss interaction potential $\Psi_{N_v} (x,y)$ between particles with coordinates $x$ and $y$ reads~\cite{KKD20,KD22}:
\begin{equation}\label{0d2}
	\Psi_{N_v} (x, y) = - \frac{J_1}{N_v} + J_2 \sum_{l =1}^{N_v}
	I_{\Delta_l}(x) I_{\Delta_l}(y).
\end{equation}
The first term in~\eqref{0d2} represents a global mean-field–type attraction with the coupling constant $J_1 > 0$ acting between any pair of particles in the system. The second term describes a local repulsion between particles located in the same cell $\Delta_l $ characterized by the parameter $J_2 > 0$.
The indicator function $I_{\Delta_l}(x)$ is defined as
\begin{equation}	\label{0d3}
	I_{\Delta_l}(x) = \left\{
	\begin{array}{ll}
		1, & x\in \Delta_l , \\
		0, & x\notin \Delta_l .
	\end{array}
	\right .
\end{equation}
The interaction potential~\eqref{0d2} satisfies the stability condition~\cite{FR66,Ruelle70,HR07,DKPRS26} provided that $J_2\geq J_1>0$.

As reported in~\cite{KKD18,KKD20,KD22}, the model admits an exact solution in the thermodynamic limit and, at sufficiently low temperatures, exhibits an infinite sequence of first-order phase transitions between phases of increasing density. Each phase transition line terminates at its own critical point, resulting in an infinite set of critical points.

The paper is organized as follows. 
In Sect.~\ref{sec:eos} we briefly reproduce the asymptotic solution for the grand partition function and the derivation of explicit expressions for the chemical potential, the particle number density and the equation of state. In Sect.~\ref{sec:respons_functions}, the explicit expressions and plots for thermodynamic response functions are presented as well as a comparison with the van der Waals model, Lennard-Jones systems, and models with soft and penetrable effective interactions. Note that this comparison concerns the thermodynamic behavior of the response functions rather than their microscopic origin. 
The subcritical phase diagrams of the CFM are displayed for reference in Appendix~\ref{sec:app-a}. Technical details of the calculations are given in Appendices~\ref{sec:app-b}~--~\ref{sec:app-d}.

\section{\label{sec:eos} Equation of state and known relations}

In this section, we briefly outline the main results for the CFM following \cite{KKD20} and \cite{RDKPS26a}.
We employ the dimensionless thermodynamic variables commonly used in the statistical-mechanical description of fluids~\cite{HansenMcDonald13}: the temperature $T^*=k_{\mathrm{B}}T/J_1$, pressure $P^*=Pv/J_1$, chemical potential $\mu^*=\mu/J_1$, and particle number density $\rho^*={v \langle N \rangle}/{V} ={\langle N \rangle}/{N_v}$, where $\langle\cdots\rangle$ denotes the grand-canonical ensemble average. We also introduce the dimensionless cell volume $v^*=v/\Lambda_J^3$, where $\Lambda_J=(2\pi\hbar^2/mJ_1)^{1/2}$, and the dimensionless ratio $f=J_2/J_1$ of the repulsion and attraction couplings. The grand partition function of the system at temperature $T$ and chemical potential $\mu$ is given by

\begin{equation}\label{1d4}
	\Xi=\sum_{N=0}^\infty \frac{\zeta^N}{N!} \int \limits_V \mathrm dx_1 ...
	\int \limits_V \mathrm dx_N \exp\left[ -\frac{\beta}{2} \sum_{x,y\in\gamma_N} \Phi_{N_v} (x,y)\right],
\end{equation}
where $\zeta =\exp(\beta\mu)/\Lambda^3$ is the activity, $\beta=(k_BT)^{-1}$ is the inverse temperature, $\Lambda= (2\pi\beta\hbar^2/m)^{1/2}$ is the de~Broglie thermal wavelength, $\hbar$ is the reduced Planck constant, and $m$ is the particle mass.
Also in~\eqref{1d4}, $\gamma_N=\{x_1,...,x_N\}$ is a configuration of $N$ particles in the volume $V$, and $x_i$ denotes the space position of the $i$-th particle.
The two-particle interaction potential $\Phi_{N_v}(x, y)$ is given in \eqref{0d2}.

In the thermodynamic limit $N_v \to \infty$ implying $V \to \infty$ with $V/N_v=v=const$, the grand partition function has the asymptotic form~\cite{KKD20,KD22,RDKPS26a}
\begin{equation}
	\label{eq:Xi1}
	\Xi(T^*,\mu^*)\propto\exp \left[ N_v E(T^*, \mu^*;\bar{z}_{\rm max})\right],
\end{equation}
where we omit a negligible constant factor.
The function $E=E(T^*,\mu^*;z)$ is given by
\begin{equation}
	\label{1d6}
	E(T^*,\mu^*;z)=- \frac{T^*}{ 2 }\left(z-\frac{\mu^*}{T^*}- \ln v^*-\frac{3}{2}\ln T^{*}\right)^2 + \ln K_0(T^*;z),
\end{equation}
where $K_0(T^*; z)$ is the $0$-th member of the family of special functions
\begin{eqnarray}\label{def:Kj}
	K_j(T^*;z)=\sum_{n=0}^{\infty} \frac{n^j}{n!} \exp\left(zn- \frac{f}{2T^*}n^2\right).
\end{eqnarray}

The argument $\bar{z}_{\rm max}=\bar{z}_{\rm max}(T^*,\mu^*)$ of the function $E$ in \eqref{eq:Xi1} appears as the position of its global maximum satisfying simultaneously the couple of conditions
\begin{eqnarray}
	\label{eq:max_cond}
	\frac{\partial E(T^*,\mu^*;z)}{\partial z}\bigg|_{z =\bar{z}_{\rm max}}=0
	\qquad\mbox{and}\qquad
	\frac{\partial^2 E(T^*,\mu^*;z)}{\partial z^2}\bigg|_{z =\bar{z}_{\rm max}}<0
\end{eqnarray}
along with the requirement
\begin{equation}\label{REQ}
	E(\bar{z}_{\rm max})>E(z),\qquad\forall z\,\in(-\infty,\infty)\,,
\end{equation}
related to the application of Laplace's method in evaluation of the grand partition function in the thermodynamic limit~\cite{KD22}.

In the present research, we consider only supercritical region ($T>T_c$). In the region above the critical points, any solution $\bar{z}$ to the system of equations~\eqref{eq:max_cond} leads to the global maximum of the function $E$ in \eqref{eq:Xi1}. 

The relation between $\mu^*$ and $\bar{z}_{\rm max}$ resulting from the first, extremum condition in~\eqref{eq:max_cond}, is given by
\begin{equation}\label{1d10}
	\mu^* (T^*;\bar{z}_{\rm max})=T^*\bar{z}_{\rm max} -  \frac{K_1(T^*;\bar{z}_{\rm max})}{K_0(T^*;\bar{z}_{\rm max})} -T^* \ln v^*-\frac{3}{2} T^* \ln T^{*}.
\end{equation}

In this case, the standard thermodynamic relation for the average number of particles $\langle N \rangle=\beta^{-1}\partial\ln\Xi/\partial\mu$ can be applied, and this yields the particle number density $\rho^*=\langle N \rangle / N_v$ as
\begin{equation}\label{1d11}
	\rho^*(T^*;\bar{z}_{\rm max})=\frac{K_1(T^*;\bar{z}_{\rm max})}{K_0(T^*;\bar{z}_{\rm max})}\equiv M_1(T^*;\bar{z}_{\rm max})\,,
\end{equation}
where we identify the ratio $K_1/K_0$ in the middle of \eqref{1d11} with the first moment of the discrete Gauss-Poisson distribution implied in \eqref{def:Kj} (cf. \cite[(25)]{KD22} and \cite{DS24arxiv}).

In other words, the definition $\rho^*=\langle N \rangle/N_v$ means that $\rho^*$ has equivalently the meaning of the mean cell occupancy. This is a crucial physical interpretation of $\rho^*$ for the CFM.

We substitute the chemical potential $\mu^*$ from \eqref{1d10}, along with \eqref{1d11}, into~\eqref{1d6}. This eliminates the explicit dependence on $\mu^*$ from~\eqref{1d6}, which transform into
\begin{equation}\label{EF}
	E(T^*;\bar{z}_{\rm max})=\ln K_0(T^*;\bar{z}_{\rm max})-\frac{1} {2T^*}\left[M_1(T^*;\bar{z}_{\rm max})\right]^2.
\end{equation}
The grand partition function \eqref{eq:Xi1} becomes thus
\begin{equation}\label{GGP}
	\Xi(T^*,\mu^*)\propto\exp\left[N_v E\left( T^*;\bar{z}_{\rm max}(T^*,\mu^*) \right)\right].
\end{equation}

Using the grand partition function \eqref{GGP} in the standard thermodynamic relation $P V=k_{\rm B} T \ln \Xi$, the equation of state can be written in dimensionless form as
\begin{equation}\label{1d16}
	P^*=T^* E(T^*;\bar{z}_{\rm max}).
\end{equation}
Combined, equations~\eqref{1d11} and~\eqref{1d16} yield a parametric representation of the equation of state in pressure–-temperature–-density coordinates with $\bar{z}_{\rm max}$ serving as a parameter. The chemical potential~\eqref{1d10} can be expressed analogously as a function of the density and temperature. Except for the chemical potential~\eqref{1d10}, neither the particle number density~\eqref{1d11} nor the pressure~\eqref{1d16} depends on the cell volume $v^*$.
\begin{figure}[t]
	\centering
	\includegraphics[width=0.465\textwidth]{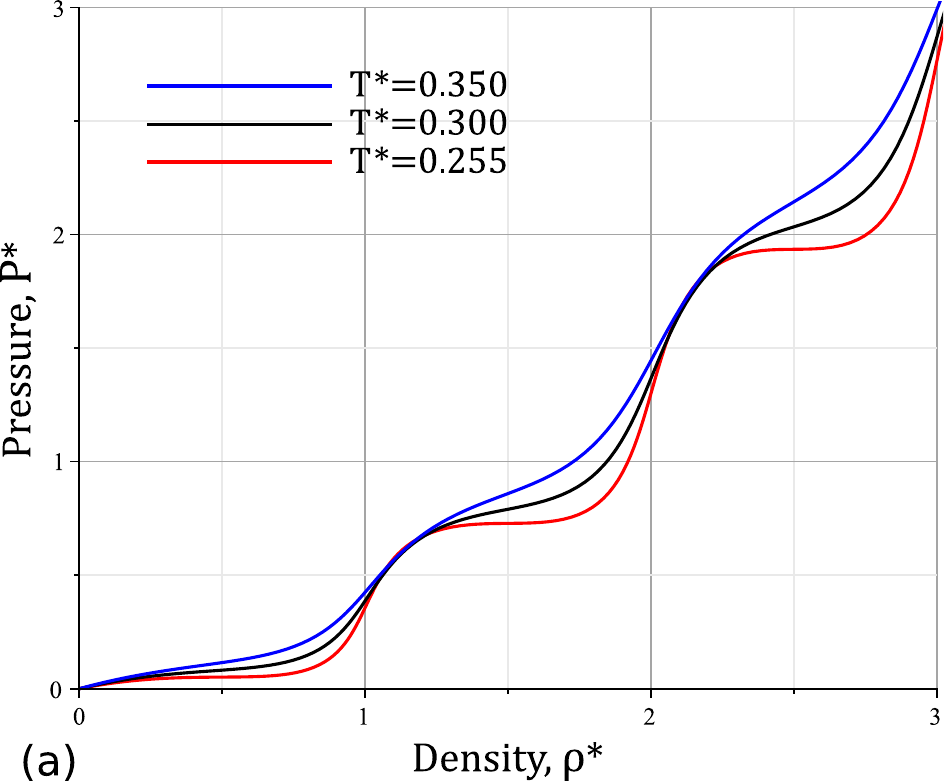}
	\includegraphics[width=0.465\textwidth]{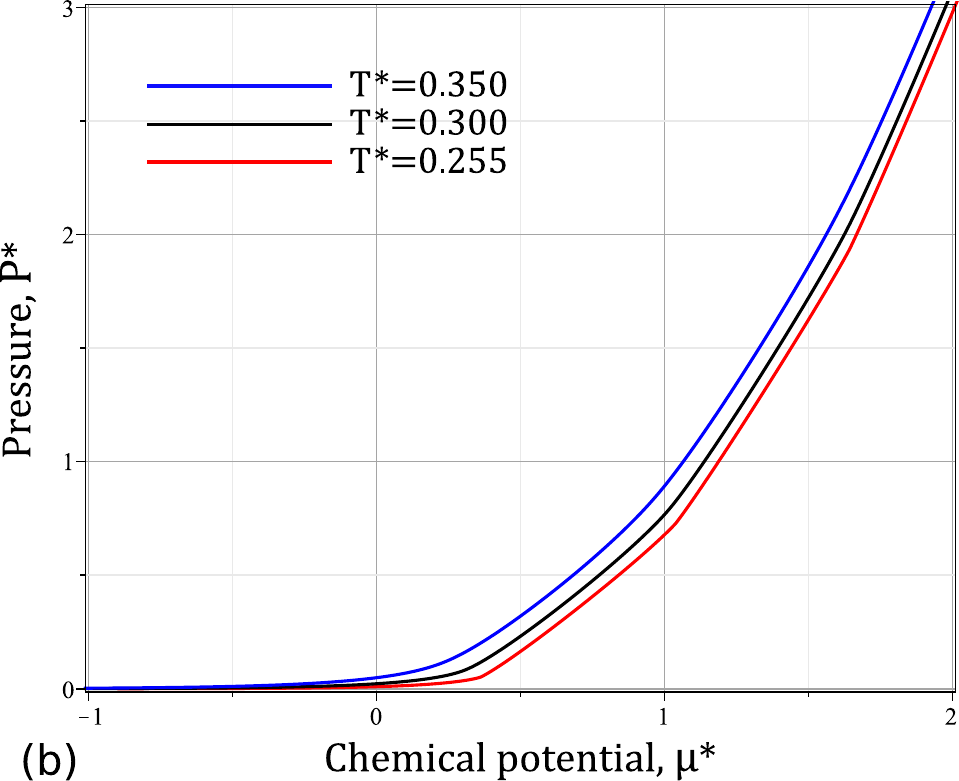}
	\captionsetup{width=0.9\textwidth}
	\caption{The pressure $P^*$ as a function of the particle number density $\rho^*$ (a) and the chemical potential $\mu^*$ (b) for several temperatures $T^*$ in the supercritical region at $f=1.5$; $\mu^*=\mu^*(v^*=5.0)$.
		}
	\label{fig2}
\end{figure}

The pressure isotherms obtained from the parametric equation of state are shown in Fig.~\ref{fig2} as functions of the particle number density and the chemical potential for several supercritical temperatures. The ratio of the repulsive to attractive coupling strengths is fixed at $f=1.5$, while the chemical potential $\mu^*$ is evaluated at $v^*=5.0$, i.e., $\mu^*=\mu^*(v^*=5.0)$. 
Within the density interval shown in Fig.~\ref{fig2}a, the corresponding subcritical region contains three consecutive first-order phase transitions, as presented in Appendix~\ref{sec:app-a}, Fig.~\ref{fig0}. The associated critical points are listed in Table~\ref{tab1} of Appendix~\ref{sec:app-a}. Their critical temperatures are $T_c^{*(k)}=0.251285$, $0.253109$, and $0.253838$, respectively. Since all temperatures considered in Fig.~\ref{fig2} exceed the highest critical temperature of this sequence, the system remains in the supercritical regime throughout the displayed range. Accordingly, each pressure isotherm is a single-valued and smooth function of both $\rho^*$ and $\mu^*$, with no signatures of phase coexistence or metastability~\cite{LJ24,OYL24}.

Figure~\ref{fig2}(a) shows that the pressure increases monotonically with the particle number density for all investigated temperatures, in agreement with the requirement of mechanical stability. As expected, the pressure becomes larger both with increasing density and with increasing temperature. At relatively low densities, the pressure grows slowly, whereas at higher densities the influence of the repulsive interaction becomes increasingly pronounced, leading to a steeper increase of the pressure.

The pressure also increases monotonically with the chemical potential, as shown in Fig.~\ref{fig2}(b). This behavior follows directly from the thermodynamic identity $(\partial P / \partial \mu)_T = \rho $, which guarantees a positive slope since the particle number density is always positive. Consequently, the pressure--chemical potential representation provides an alternative but equivalent description of the equation of state.

The smooth character of the isotherms shown in Fig.~\ref{fig2} reflects the absence of thermodynamic singularities in the supercritical region~\cite{LJ24}. Nevertheless, as will be demonstrated in the following sections, the derivatives of the equation of state exhibit pronounced anomalies upon approaching the critical temperatures.

\section{\label{sec:respons_functions}Thermodynamic response functions}

For a simple one-component system, the thermodynamic behavior is fully determined by three independent response functions~\cite[Chapter~7]{Callen85}. A convenient choice is the isothermal compressibility, the thermal pressure coefficient and the isochoric heat capacity. All remaining coefficients, such as the isobaric heat capacity, and the thermal expansion coefficient, follow from these through general thermodynamic identities~\cite[Chapter~7]{Callen85}.

\subsection{\label{sec:KT}Isothermal compressibility }
The isothermal compressibility is defined by~\cite[(18)]{SM21}
\begin{equation}
	\label{def:isotherm_compres}
	\kappa_T = -\frac{1}{V}\left(\frac{\partial V}{\partial P}\right)_{T,  N} .
\end{equation}
Let us rewrite $\kappa_T$ in equivalent form using $\rho = N/V$:
\begin{equation}\label{KT_equiv}
	\kappa_T	 =  \frac{1}{\rho} \left(\frac{\partial \rho}{\partial P} \right)_{T} = \frac{1}{\rho} \frac{\left(\partial \rho / \partial \mu\right)_{T}}
	{\left(\partial P / \partial \mu\right)_{T}}. 
\end{equation}
The condition of constant $ N $ is omitted in the last expression of \eqref{KT_equiv} since we have explicit dependence on temperature and chemical potential for both the particle number density $\rho^* = \rho^*(T^*; \bar{z}_{\rm max}(T^*,\mu^*))$, Eq.~\eqref{1d11}, and the pressure $P^* = P^*(T^*; \bar{z}_{\rm max}(T^*,\mu^*))$, Eq.~\eqref{1d16}.
Applying the Gibbs--Duhem equation
\begin{equation}
	 N {\rm d}\mu = -S{\rm d}T + V{\rm d}P,
\end{equation}
at $T = const$ one has 
\begin{equation*}
	{\rm d} P = \rho {\rm d} \mu
\end{equation*}
or
\begin{equation}
	\label{eq:rho_dpdm}
	\rho = \left(\frac{\partial P}{\partial \mu}\right)_{T}.
\end{equation}
Substituting this into the last expression for $\kappa_T$ in \eqref{KT_equiv}, we obtain
\begin{equation}
	\kappa_T = \frac{1}{\rho^2} \left(\frac{\partial \rho}{\partial \mu}\right)_{T} .
\end{equation}

Let us introduce the dimensionless isothermal compressibility
\begin{equation}
	\kappa^*_T \equiv \frac{J_1 \kappa_T}{v}.
\end{equation}
The quantity $\kappa^*_T$ is of order unity, except at the critical point itself, where it is divergent. It is expressed in terms of the chemical potential $\mu^*$ and particle number density $\rho^*$ via
\begin{equation}
	\label{eq:kappa_star_m1}
	\kappa^*_T  =  \frac{1}{{\rho^*}^2} \left(\frac{\partial \rho^*}{\partial \mu^{*}}\right)_{T^*} .
\end{equation}

We now use Eq.~\eqref{eq:kappa_star_m1} to explicitly calculate $\kappa^*_T$ (see Appendix~\ref{sec:app-b}), with the result expressed as a function of temperature and $\bar{z}_{\rm max}$ as follows
\begin{equation}\label{eq:kappa_exp}
	\kappa^*_T  =  \frac{1}{{\rho^*}^2 (T^*;\bar{z}_{\rm max}) }  \frac{	M_2(T^*;\bar{z}_{\rm max}) }{ T^* - M_2(T^*;\bar{z}_{\rm max}) },
\end{equation}
where we identify 
\begin{equation}\label{eq:m2}
	M_2(T^*;\bar{z}_{\rm max})  =  \frac{K_2(T^*;\bar{z}_{\rm max})}{K_0(T^*;\bar{z}_{\rm max})} - [M_1(T^*;\bar{z}_{\rm max})]^2 
\end{equation}
with the second moment of the discrete Gauss-Poisson distribution implied in \eqref{def:Kj} (cf. \cite[(25)]{KD22} and \cite{DS24arxiv}). 
\begin{figure}[t]
	\centering
	\includegraphics[width=0.465\textwidth]{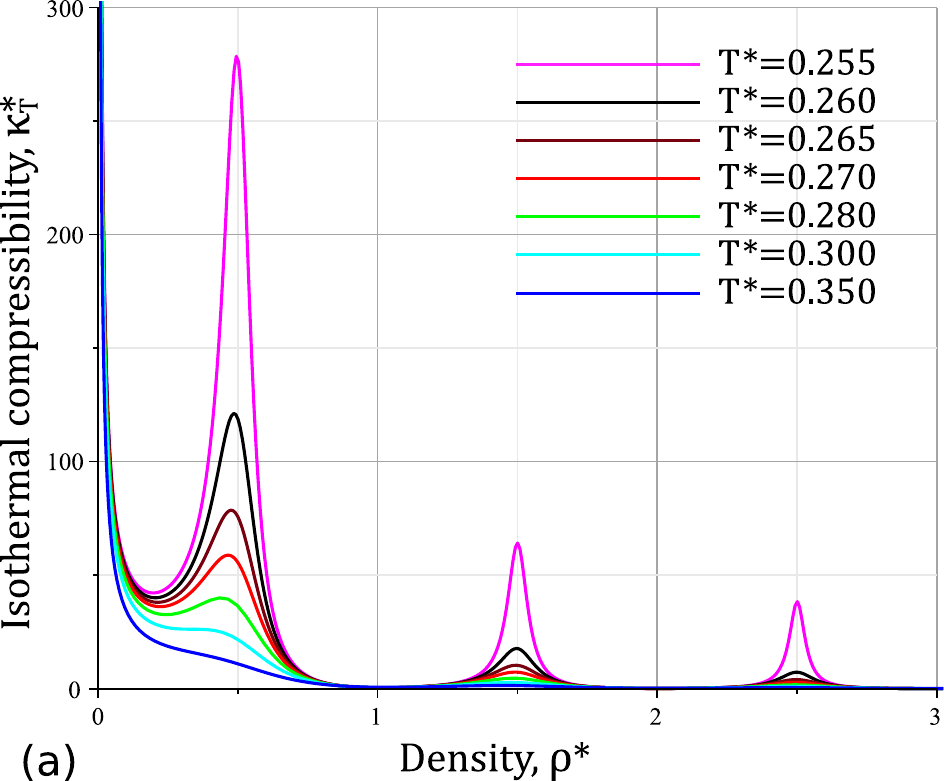}
	\includegraphics[width=0.465\textwidth]{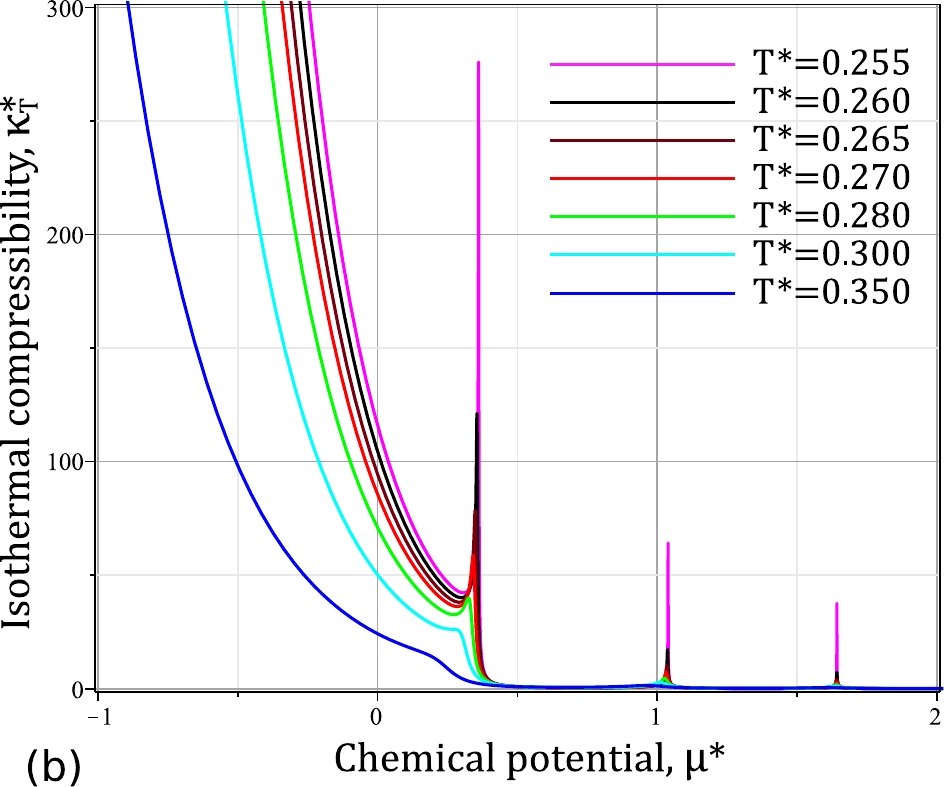}
	\captionsetup{width=0.9\textwidth}
		\caption{The isothermal compressibility $\kappa^*_T$ as a function of the particle number density $\rho^*$ (a) and the chemical potential $\mu^*$ (b) at different supercritical values of temperature $T^*$. Parameters taken as $f=1.5$, $v^*= 5.0$. }
		\label{fig3}
\end{figure}
Figure~\ref{fig3} illustrates the dependence of the compressibility $\kappa^*_T$ on the particle number density $\rho^*$ and the chemical potential $\mu^*$ for various values of supercritical temperature $T^*$. The isothermal compressibility remains strictly positive throughout the thermodynamically stable region, in agreement with the general stability criterion requiring the pressure to increase monotonically with density. Consequently, the positive values of $\kappa_T^*$ confirm that all calculated equilibrium states satisfy the condition of mechanical stability.

As the temperature approaches a critical value from above, the isothermal compressibility develops increasingly pronounced maxima centered at densities close to the corresponding critical densities, $\rho^*_c\approx0.5,1.5,2.5,\ldots$ (see Table~\ref{tab1} in Appendix~\ref{sec:app-a}). 
Within the present mean-field description, this behavior is a direct consequence of the analytical structure of the equation for $\kappa_T^*$. According to Eq.~\eqref{eq:kappa_exp}, the denominator contains the factor $T^*-M_2(T^*;\bar z_{\rm max})$. At the critical points this quantity vanishes, leading to the divergence of the isothermal compressibility. 

Although the present model does not explicitly account for microscopic spatial fluctuations, it correctly reproduces the thermodynamic signature of criticality through the singular behavior of the equation of state. In this sense, the divergence of $\kappa_T^*$ reflects the loss of rigidity of the equilibrium thermodynamic state: near the critical point the system becomes increasingly susceptible to external compression, so that an arbitrarily small change in pressure is sufficient to produce a macroscopic change in the equilibrium density.

The repeated appearance of compressibility maxima at densities close to critical values reflects the distinctive feature of the present cell fluid model, namely the existence of an infinite hierarchy of fluid-fluid critical points. Each maximum is associated with one member of this hierarchy and evolves into a divergence at the corresponding critical temperature. Thus, the compressibility provides a direct indication of the successive critical states predicted by the model.

It is instructive to compare the behavior of $\kappa_T^*$ obtained for the present CFM with that of other fluid models. For the van der Waals fluid, the isothermal compressibility exhibits a pronounced maximum in the supercritical region, which becomes increasingly sharp on approaching the critical point~\cite{Johnston14,LJ24,WAP24}. The corresponding line of compressibility maxima terminates at the critical point. Similar behavior is observed in the CFM with Morse interaction~\cite{DKRP24,KPD18}, Lennard--Jones~\cite{YS13,MM14} and square-well-potential systems~{\cite{SN11}}, Gaussian core model~\cite{LS19} and real fluids~\cite{VRMBOA11,WAP24} in the supercritical vicinity of the vapor--liquid critical point. These results indicate that the enhancement of $\kappa_T^*$ observed in the present model is consistent with the general thermodynamic behavior associated with fluid criticality.
The common occurrence of a strongly enhanced isothermal compressibility near criticality implies that this feature is a robust thermodynamic signature of critical behavior, while the repeated sequence of maxima found in the present paper reflects the distinctive phase structure of the CFM.

\subsection{\label{sec:BV}Thermal pressure coefficient}
The thermal pressure coefficient is defined by~\cite[(17)]{SM21}
\begin{equation}
	\label{def:therm_pres_coef}
	\beta_V = \left( \frac{\partial P}{\partial T} \right)_{V,N}.
\end{equation}
Using \eqref{1d16} and \eqref{1d11} we rewrite $\beta_V$ in a form suitable for the equation of state $P^* = P^*(T^*; \bar{z}_{\rm max}(T^*,\mu^*))$, in~\eqref{1d16}, transformed by the method of Jacobian determinants with respect to the constant volume and number of particles (see also (17) from~\cite{SM21}):
\begin{equation}
	\beta_V = \left(\frac{\partial P}{\partial T}\right)_{V,\mu} 
	- \left(\frac{\partial P}{\partial \mu}\right)_{T,V} 
	\left(\frac{\partial \rho^*}{\partial T}\right)_{V,\mu}
	\left(\frac{\partial \rho^*}{\partial \mu}\right)^{-1}_{T,V}.
\end{equation}
Here, the first contribution to $\beta_V$ is essentially the entropy per cell, $S/N_v = (\partial P / \partial T)_{V,\mu}$. 

Let us introduce the dimensionless thermal pressure coefficient via
\begin{equation}
	\beta^*_V = \frac{v}{k_{\rm B}}\beta_V.
\end{equation}
It is expressed in terms of dimensionless quantities as
\begin{equation}
	\label{eq:beta_star_m}
	\beta^*_V = \left(\frac{\partial P^*}{\partial T^*}\right)_{\mu^*} 
	- \left(\frac{\partial P^*}{\partial \mu^*}\right)_{T^*} 
	\left(\frac{\partial \rho^*}{\partial T^*}\right)_{\mu^*}
	\left(\frac{\partial \rho^*}{\partial \mu^*}\right)^{-1}_{T^*} .
\end{equation}
The explicit equation of the thermal pressure coefficient for the cell fluid model with the Curie-Weiss interaction is derived in Appendix~\ref{sec:app-c} and given by:
\begin{align}\label{eq:beta_ex}
	\beta^*_V & = \ln K_0(T^*;\bar{z}_{\rm max}) + \frac{f}{2 T^*} \frac{K_2(T^*;\bar{z}_{\rm max})}{K_0(T^*;\bar{z}_{\rm max})}  - M_1(T^*;\bar{z}_{\rm max}) \left(\ln v^* + \frac{3}{2} \ln T^*\right) \nonumber  \\ 
	& - \frac{f M_1(T^*;\bar{z}_{\rm max})}{ 2 T^* M_2(T^*;\bar{z}_{\rm max})}\left( \frac{K_3(T^*;\bar{z}_{\rm max})}{K_0(T^*;\bar{z}_{\rm max})} - M_1(T^*;\bar{z}_{\rm max}) \frac{K_2(T^*;\bar{z}_{\rm max})}{K_0(T^*;\bar{z}_{\rm max})} \right).
\end{align}

\begin{figure}[t]
	\centering
	\includegraphics[width=0.465\textwidth]{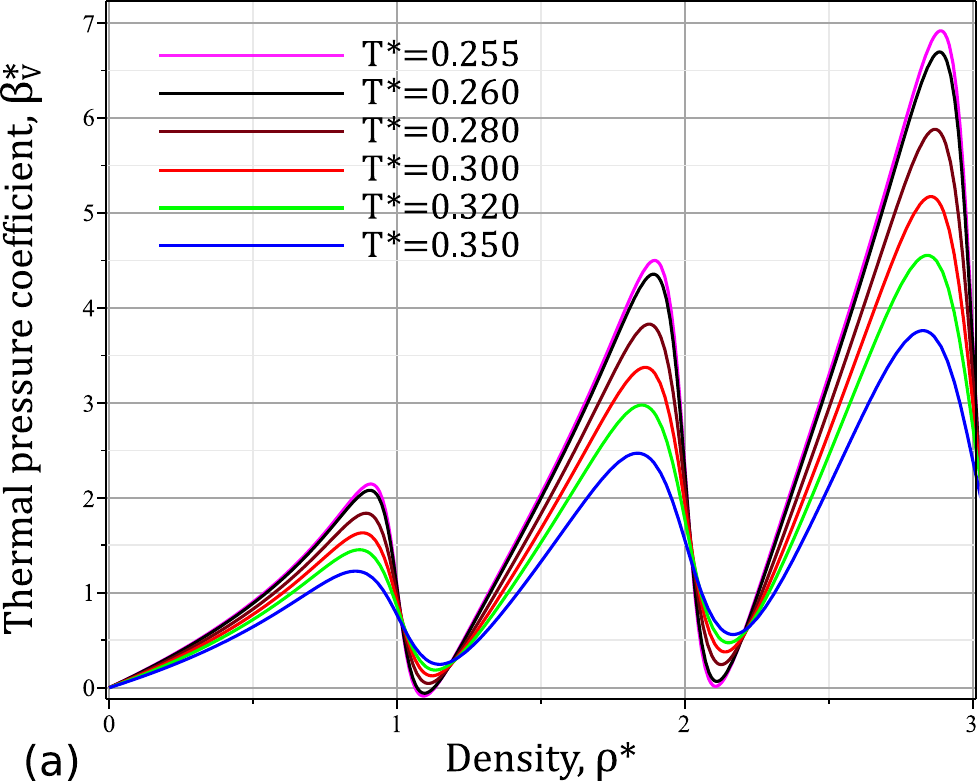}
	\includegraphics[width=0.465\textwidth]{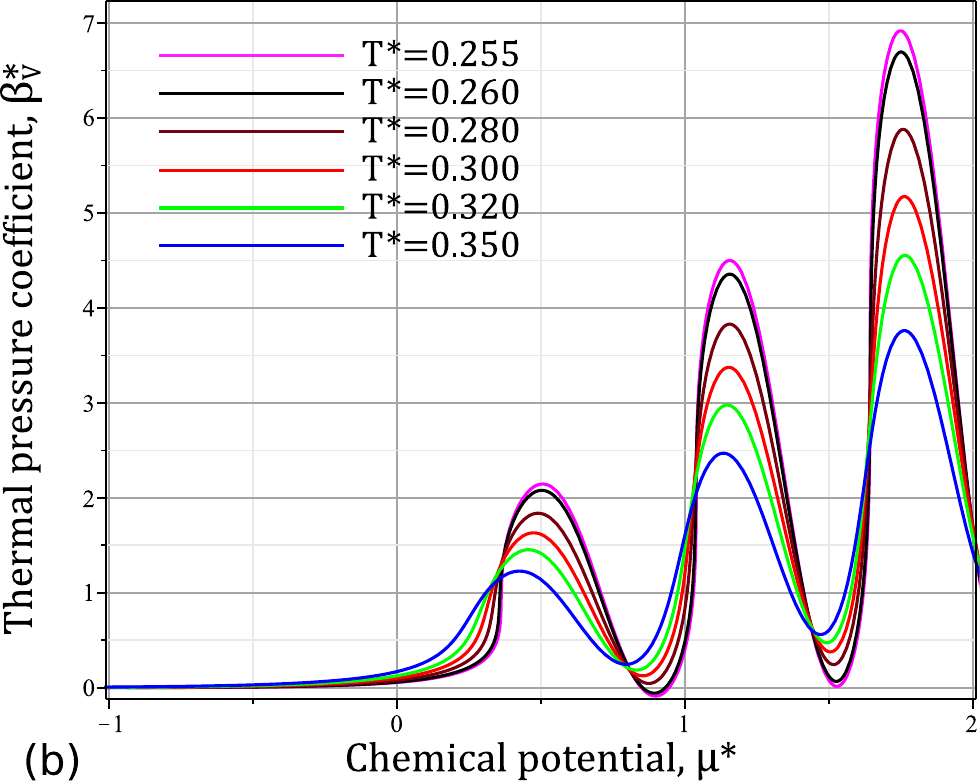}
	\captionsetup{width=0.9\textwidth}
	\caption{The thermal pressure coefficient $\beta^*_V$ as a function of the particle number density $\rho^*$ (a) and the chemical potential $\mu^*$ (b) at different values of temperature $T^*$ in the supercritical region. Parameters taken as $f=1.5$, $v^*= 5.0$.}
	\label{fig4}
\end{figure}

Figure~\ref{fig4} shows the dependence of the thermal pressure coefficient $\beta^*_V$ on the particle number density $\rho^*$ and the chemical potential $\mu^*$ for six different values of supercritical temperature $T^*$.

The thermal pressure coefficient measures the change in pressure produced by a temperature variation at fixed volume. In other words, it quantifies how strongly the system responds to heating when the density is kept constant. Since it is determined by the temperature dependence of both the entropy and the equilibrium quantity $\bar{z}_{\rm max}$, its behavior reflects the competition between the thermal contribution and the interaction contribution to the pressure.

The thermal pressure coefficient exhibits a qualitatively different behavior from the isothermal compressibility. Although it shows a pronounced non-monotonic dependence on density, its extrema are not sharp and are not located at the critical densities, but rather occur approximately midway between successive critical states. Moreover, unlike the compressibility, $\beta_V^*$ remains finite at critical points. This behavior is consistent with the analytical expression~\eqref{eq:beta_ex}, which contains no denominator that vanishes under the criticality condition.

The oscillatory character of $\beta_V^*$ indicates that the thermal response of the pressure is governed by the detailed evolution of the equation of state over the entire density range rather than solely by the vicinity of the critical points. Successive maxima and minima arise as the system passes through the sequence of equilibrium states characteristic of the present model, reflecting the varying balance between thermal effects and intermolecular interactions.

An even more remarkable feature is the appearance of negative values of the thermal pressure coefficient at relatively low temperatures and densities around $\rho^*\approx1.0$--$1.2$. A negative value of $\beta_V^*$ means that, at constant volume, an increase in temperature leads to a decrease in pressure. Although this behavior is counterintuitive from the perspective of ordinary fluids, it is thermodynamically admissible and indicates that, in this density interval, the interaction contribution to the pressure changes with temperature more rapidly than the direct thermal contribution. The resulting competition between these two mechanisms produces an anomalous thermal response without violating the stability conditions of the equilibrium state.

Finally, the thermal pressure coefficient exhibits a much stronger temperature dependence than that observed for the cell fluid model with the Morse interaction~\cite{DKRP24}, the van der Waals fluid~\cite{Johnston14}, and the Lennard-Jones fluid~\cite{YS13,MM14}. In the latter cases, $\beta_V^*$ was found to increase almost monotonically with density and depended only weakly on temperature. 

Interesting similarities are also found when the comparison is extended to soft-core and ultrasoft interactions. For the Gaussian-core model, the bounded repulsive potential permits particle overlap and produces a number of anomalous thermodynamic properties, including regions in which the thermal pressure coefficient becomes negative. Molecular simulations of the Gaussian-core and related double-Gaussian-core models have explicitly demonstrated negative thermal pressure coefficient together with anomalies in the thermal expansion coefficient and isothermal compressibility~\cite{LS19,MS11}. This provides a useful qualitative comparison with the negative values of $\beta_V^*$ obtained in the present CFM. 

An analysis of our results along with those for other models with different interactions indicates that the thermal pressure coefficient is considerably more sensitive to the microscopic form of the interaction than the isothermal compressibility. Whereas the latter exhibits a common critical enhancement expected for a mechanically susceptible fluid, $\beta_V$ can remain finite at criticality while displaying qualitatively different density and temperature dependencies, including monotonic, anomalous, or non-monotonic behavior depending on the interaction potential. In the present CFM, the pronounced temperature dependence, oscillatory density dependence, and occurrence of negative values of $\beta_V^*$ demonstrate that this response function provides a particularly sensitive probe of the competition between the thermal and interaction contributions to the equation of state.


\subsection{\label{sec:CV}Isochoric heat capacity}
The isochoric heat capacity in canonical ensemble is defined by~\cite[(13)]{SM21}
\begin{equation}\label{eq:cv_def}
	C_V = \left(\frac{\partial U}{\partial T}\right)_{V,N},
\end{equation}
where $U = \Omega + TS + PV$ is the internal energy, $\Omega = - k_B T \ln \Xi$ is the grand thermodynamic potential.
Transformed by the method of Jacobian determinants the equation \eqref{eq:cv_def} reads
\begin{equation}\label{eq:cv_trans}
	C_V = \left(\frac{\partial U}{\partial T}\right)_{V,\mu} - \left(\frac{\partial \mu}{\partial N}\right)_{T,V}\left(\frac{\partial N}{\partial T}\right)_{V,\mu} \left(\frac{\partial U}{\partial \mu}\right)_{T,V}.
\end{equation}
Using the definitions:
\begin{equation}\label{eq:1}
	S = - \left(\frac{\partial \Omega}{\partial T}\right)_{V,\mu}, \qquad 	N = - \left(\frac{\partial \Omega}{\partial \mu}\right)_{T,V},
\end{equation}
we can write internal energy directly as:
\begin{equation}\label{eq:u_omega}
	U = \Omega - T\left(\frac{\partial \Omega}{\partial T}\right)_{V,\mu} - \mu \left(\frac{\partial \Omega}{\partial \mu}\right)_{T,V}.
\end{equation}
Combining the equations \eqref{eq:cv_trans} and \eqref{eq:u_omega} we rewrite $C_V$ into a form suitable for the equation of state $P^* = P^*(T^*; \bar{z}_{\rm max}(T^*,\mu^*))$, see~\eqref{1d16},
\begin{equation}\label{eq:3}
		C_V = T \left( \left(\frac{\partial S}{\partial T}\right)_{V,\mu} - \left(\frac{\partial \mu}{\partial N}\right)_{T,V}\left(\frac{\partial N}{\partial T}\right)_{V,\mu} \left(\frac{\partial S}{\partial \mu}\right)_{T,V} \right).
\end{equation}

We introduce the dimensionless isochoric heat capacity per particle by
\begin{equation}\label{eq:4}
	C^*_V = \frac{C_V}{k_{\rm B} \langle N \rangle},
\end{equation}
as well as the dimensionless isochoric heat capacity per cell:
\begin{equation}\label{eq:5}
	c^*_V = \frac{C_V}{k_{\rm B} N_v}.
\end{equation}
The connection between $c^*_V$ and $C^*_V$ is:
\begin{equation}\label{eq:5a}
	c^*_V = \rho^* C^*_V.
\end{equation}
\begin{figure}[t]
	\centering
	\includegraphics[width=0.465\textwidth]{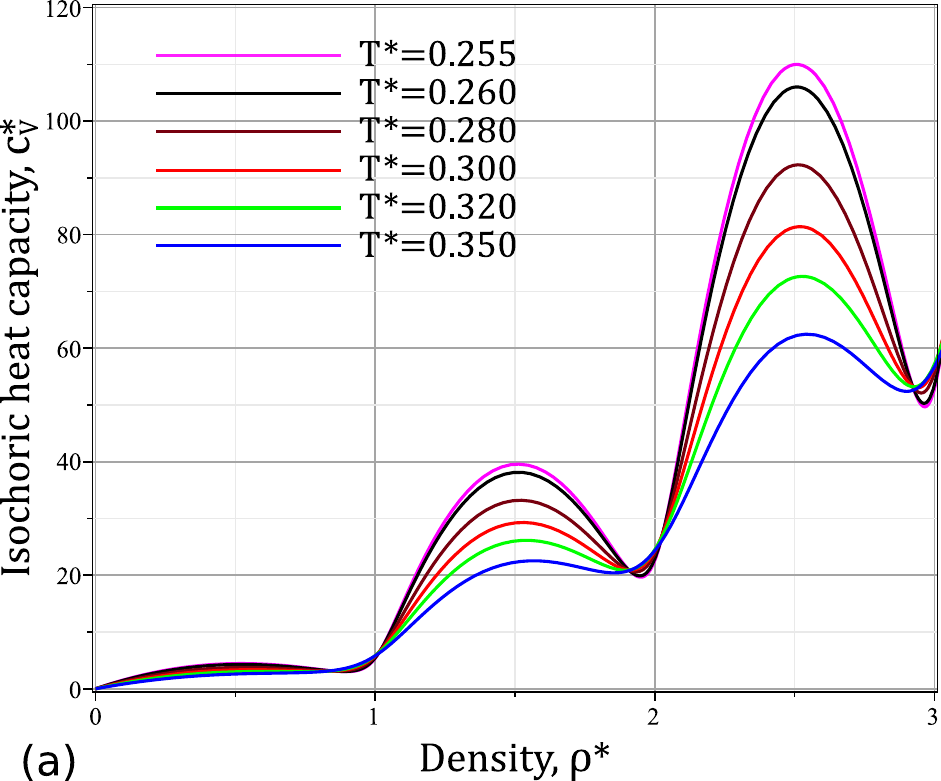}
	\includegraphics[width=0.465\textwidth]{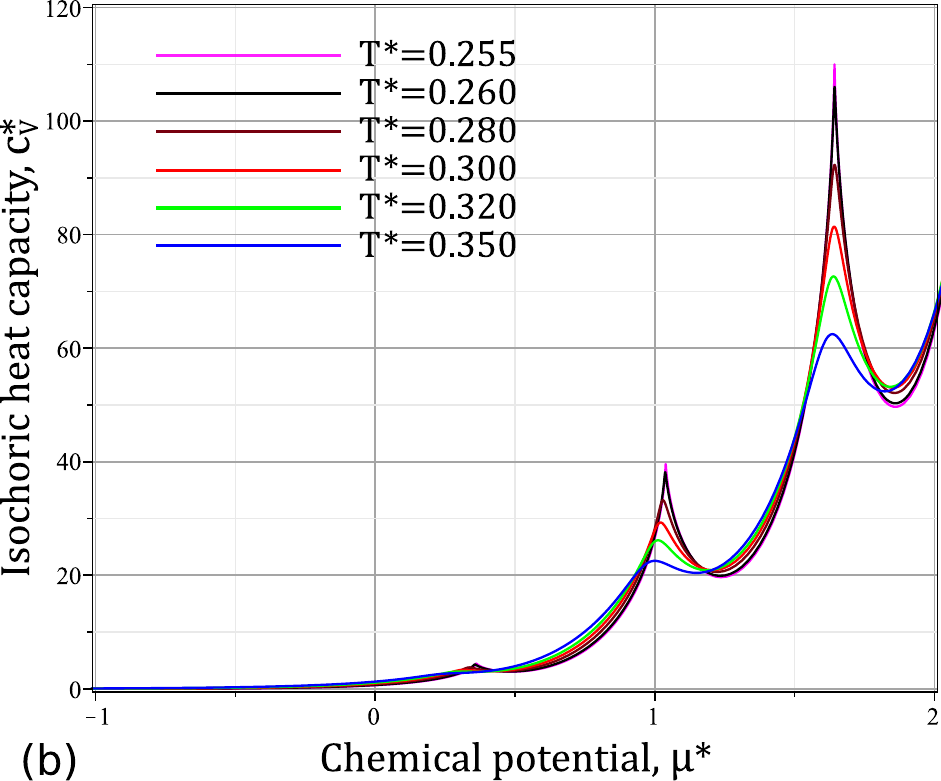}
	\captionsetup{width=0.9\textwidth}
	\caption{The isochoric heat capacity per cell $c^*_V$ as a function of the particle number density $\rho^*$ (a) and the chemical potential $\mu^*$ (b) at different values of temperature $T^*$ in the supercritical region. Parameters taken as $f=1.5$, $v^*= 5.0$.}
	\label{fig5}
\end{figure}
The isochoric heat capacity per cell $c^*_V$ is expressed in terms of dimensionless quantities by
\begin{equation}\label{eq:cv_cell}
	c^*_V = T^* \left( \left(\frac{\partial S^*_v}{\partial T^*}\right)_{\mu^*} - \left(\frac{\partial \mu^*}{\partial \rho^*}\right)_{T^*}\left(\frac{\partial \rho^*}{\partial T^*}\right)_{\mu^*} \left(\frac{\partial S^*_v}{\partial \mu^*}\right)_{T^*} \right).
\end{equation}
We take advantage of \cite[Sec. 7]{RDKPS26a} and \cite{RDKPS26b}, where both the entropy per particle $S^*$ and the entropy per cell $S^*_v$ of the cell fluid model with Curie-Weiss interaction were obtained in analytical form as functions of temperature and chemical potential. In particular, $S^*_v$ is given by (see (23) in \cite{RDKPS26b}):
\begin{equation}\label{eq:entropy_cell}
	S^*_v = \ln K_0(T^*;\bar{z}_{\rm max}) + \left(\frac{3}{2} - \bar{z}_{\rm max} \right) M_1(T^*;\bar{z}_{\rm max}) + \frac{f}{2 T^*} \frac{K_2(T^*;\bar{z}_{\rm max})}{K_0(T^*;\bar{z}_{\rm max})} .
\end{equation}
Using the Eqs. \eqref{1d10}, \eqref{1d11},  \eqref{1d16} and \eqref{eq:entropy_cell} in the formula \eqref{eq:cv_cell} we write the isochoric heat capacity per cell in the explicit form:
 \begin{align}\label{eq:cv_cell_ex}
 	c^*_V & = \frac{f^2}{4 T^{*2}} \left( \frac{K_4(T^*;\bar{z}_{\rm max})}{K_0(T^*;\bar{z}_{\rm max})}  - \left[ \frac{K_2(T^*;\bar{z}_{\rm max})}{K_0(T^*;\bar{z}_{\rm max})} \right]^2 \right . \nonumber \\
 	& \left . + \frac{1}{ M_2(T^*;\bar{z}_{\rm max})} \left[\frac{K_3(T^*;\bar{z}_{\rm max})}{K_0(T^*;\bar{z}_{\rm max})} - M_1(T^*;\bar{z}_{\rm max}) \frac{K_2(T^*;\bar{z}_{\rm max})}{K_0(T^*;\bar{z}_{\rm max})} \right]^2 \right).
 \end{align}
The detailed derivation for $c_V^*$ is given in Appendix~\ref{sec:app-d}.

Figure~\ref{fig5} shows the dependence of the isochoric heat capacity per cell $c^*_V$ on the  particle number density $\rho^*$ and the chemical potential $\mu^*$ for various values of supercritical temperature $T^*$.
The isochoric heat capacity remains strictly positive throughout the entire thermodynamically stable region. This behavior is consistent with the thermodynamic stability condition requiring the internal energy to increase monotonically with temperature at fixed density. Consequently, the positive values of $c_V^*$ confirm the thermal stability of the equilibrium states predicted by the model.

The density dependence of $c_V^*$ exhibits a sequence of pronounced maxima located in the vicinity of the critical densities. These maxima become increasingly well developed as the temperature approaches the corresponding critical value from above, indicating that the temperature dependence of the internal energy is strongest in the critical region. Nevertheless, unlike the isothermal compressibility, the isochoric heat capacity remains finite at every critical point.

Physically, the finite maxima indicate that, near the critical state, the equilibrium internal energy becomes more sensitive to temperature variations while preserving a regular temperature dependence. Within the present mean-field framework, this reflects the smooth evolution of the energetic properties of the equilibrium state rather than the appearance of a thermodynamic instability. Thus, the enhancement of $c_V^*$ signals an increased thermal response of the system without the development of a critical singularity.

The behavior of $c_V^*$ in the present model contrasts sharply with that found in other fluid models. In the van der Waals fluid, the isochoric heat capacity is identically equal to its ideal-gas value $\frac{3}{2}Nk_{\rm B}$ for all densities above $T_c$, with no density dependence whatsoever~\cite{Johnston14}.
In Lennard-Jones systems, molecular dynamics simulations reveal both maxima and minima of $C_V$ in the supercritical region, attributed respectively to the attractive well and the repulsive core of the potential, and $C_V$ diverges at the critical point in accord with three-dimensional Ising universality~\cite{YS13,MS13,WIH12,KP94}. In purely repulsive soft-matter models, the Gaussian core model (GCM), and GEM-4, the situation is qualitatively different: none of these systems possesses a gas–liquid critical point~\cite{PSG05,MCF07}, however $C_V$ increases sharply near the solid–liquid transition line and then decreases monotonically with increasing temperature without a minimum~\cite{MS11,MM12,MMPBG20}. The double-Gaussian core model (DGM), which recovers a gas–liquid transition~\cite{SPMG14}, exhibits maxima in $C_V$ above $T_c$~\cite{LS19}.
For square-well fluids of variable attraction range, a single pronounced $C_V$ maximum appears near the critical density whose height increases as the well narrows; within mean-field treatments the maximum remains finite at $T_c$, consistent with the present result, while full crossover equations of state recover the expected divergence~\cite{LSAS03,SBC05}.

\subsection{\label{sec:AP}Thermal expansion coefficient}

The thermal expansion coefficient is defined by~\cite[Table~III]{SM21}
\begin{equation}
	\alpha_P = \frac{1}{V}\left(\frac{\partial V}{\partial T}\right)_{P,N}.
\end{equation}
On the other hand, the thermal expansion coefficient $\alpha_P$ is not independent. It is related to the isothermal compressibility $\kappa_T$ and the thermal pressure coefficient $\beta_V$ by the ``triple product rule'',
\begin{equation}
	\label{eq:identity}
	\frac{\alpha_P}{\kappa_T \beta_V} = \frac{\alpha^*_P}{\kappa^*_T \beta^*_V} = 1,
\end{equation}
which can be deduced from the cyclic relation between $P$, $V$ and $T$ (for a detailed derivation see e.g. \cite[Sec. 4.3]{DKRP24}). The dimensionless thermal expansion coefficient $\alpha^*_P$ is defined as
\begin{equation}
	\alpha^*_P = \frac{J_1}{k_{\rm B}} \alpha_P ,
\end{equation}
and its explicit expression can be directly derived from the results \eqref{eq:kappa_exp} and \eqref{eq:beta_ex} for $\kappa^*_T$ and $\beta^*_V$ via
\begin{equation}\label{eq:ap}
	{\alpha^*_P} = {\kappa^*_T \beta^*_V}.
\end{equation}
\begin{figure}[h]
	\centering
	\includegraphics[width=0.465\textwidth]{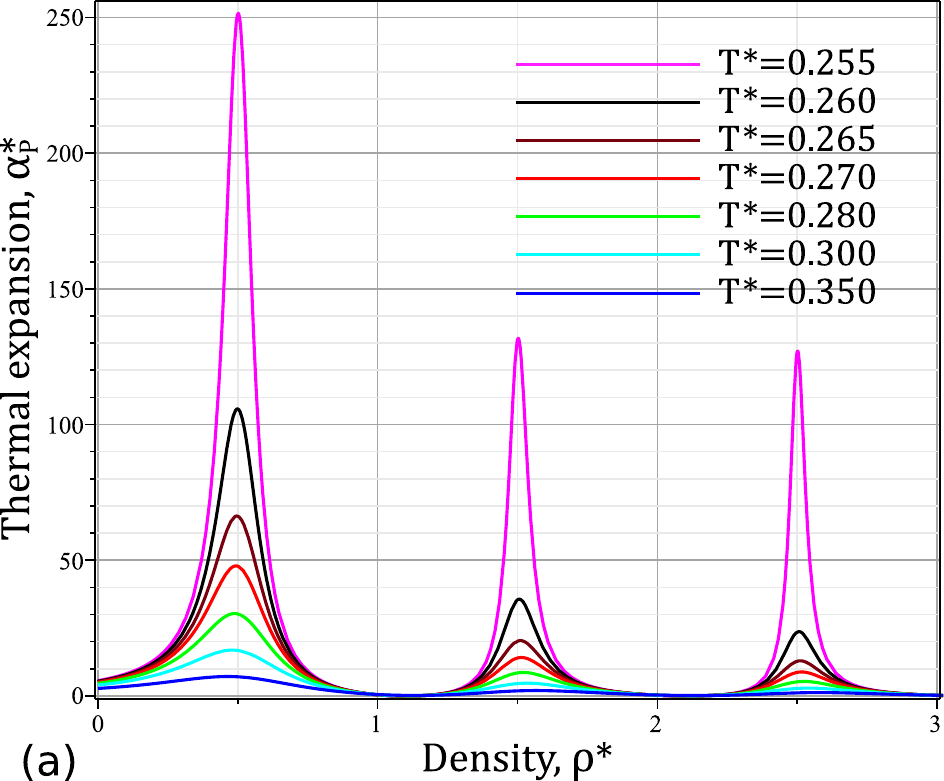}
	\includegraphics[width=0.465\textwidth]{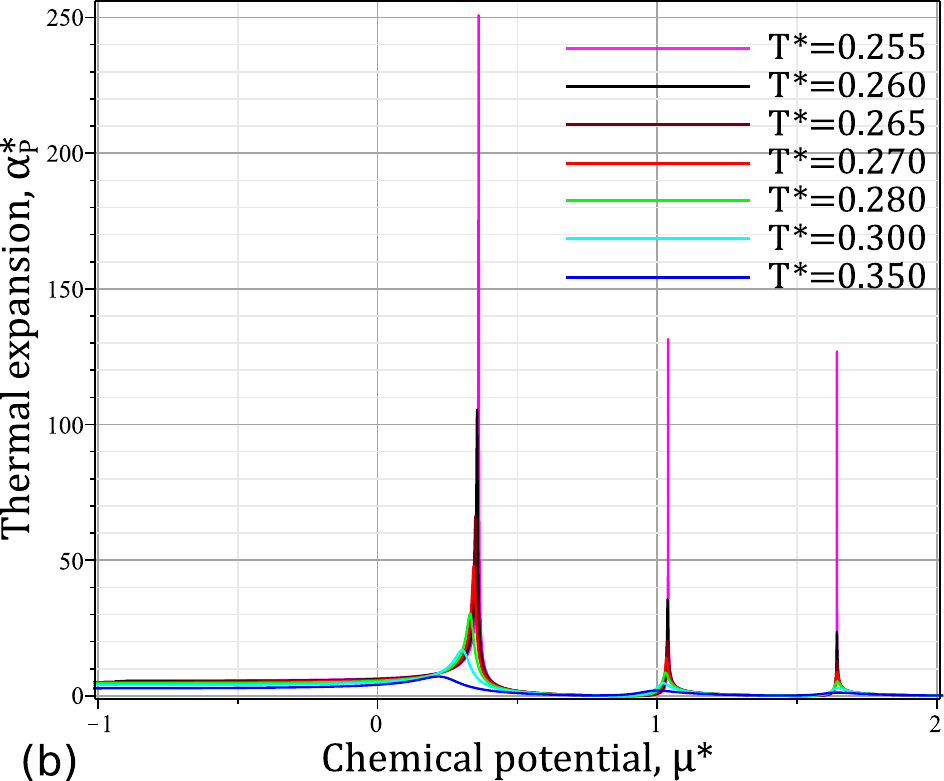}
	\captionsetup{width=0.9\textwidth}
	\caption{The thermal expansion coefficient $\alpha^*_P$ as a function of the particle number density $\rho^*$ (a) and the chemical potential $\mu^*$ (b) at different values of temperature $T^*$ in the supercritical region. Parameters taken as $f=1.5$, $v^*= 5.0$.  }
	\label{fig6}
\end{figure}
Figure~\ref{fig6} shows the dependence of the thermal expansion coefficient $\alpha^*_P$ on the  particle number density $\rho^*$ and the chemical potential $\mu^*$ for various values of supercritical temperature $T^*$.

As expected, the thermal expansion coefficient exhibits a behavior that naturally combines the isothermal compressibility and the thermal pressure coefficient. 
The most prominent feature of Fig.~\ref{fig6} is the appearance of sharp maxima of $\alpha_P^*$ in the vicinity of each critical point. In contrast to the thermal pressure coefficient, the thermal expansion coefficient diverges at critical points. This is a direct consequence of the singular behavior of the isothermal compressibility, while the thermal pressure coefficient remains finite. Therefore, the critical singularity of $\alpha_P^*$ is entirely governed by the mechanical instability encoded in $\kappa_T^*$.

Away from the critical region, the thermal expansion coefficient follows the behavior inherited from the thermal pressure coefficient $\beta_V^*$. In particular, it becomes negative at densities around $\rho^*\approx1.0$ -- $1.2$. A negative thermal expansion coefficient means that, under constant pressure, the equilibrium density increases with increasing temperature, or equivalently, that the volume decreases upon heating. This anomalous behavior originates directly from the negative values of the thermal pressure coefficient owing to the thermodynamic relation~\eqref{eq:ap}, whereas the isothermal compressibility remains strictly positive throughout the thermodynamically stable region. 

Negative thermal expansion is not unique to the present model. Experimentally, liquid water exhibits $\alpha_P<0$ below its temperature of maximum density at ambient pressure, where heating leads to an increase rather than a decrease of density~\cite{PHN16}. Negative thermal expansion has also been reported in model fluids with bounded or core-softened interactions, including the Gaussian core model~\cite{MS11,SS97}, the double-Gaussian core model~\cite{LS19}, systems with step-function potentials~\cite{Yoshimura91,KY04}, and others~\cite{BBBA05}. In such systems, the anomaly is generally associated with a competition between distinct local structures or characteristic interaction distances. In the present CFM, the relation $\alpha_P^*=\kappa_T^*\beta_V^*$ together with the strictly positive $\kappa_T^*$ establishes that the negative thermal expansion observed around $\rho^*\simeq1.0$--$1.2$ originates directly from the negative thermal pressure coefficient. Thus, although negative thermal expansion is a known feature of anomalous liquids, its occurrence here arises from the particular balance of thermal and interaction contributions in the model's equation of state.

\subsection{\label{sec:CP} Isobaric heat capacity}

The isobaric heat capacity in the grand canonical ensemble is defined by~\cite[(21)]{SM21}
\begin{equation}\label{eq:cp_def}
	C_P = \left(\frac{\partial H}{\partial T}\right)_{P,N},
\end{equation}
where $H = U + PV$ is the enthalpy.
Similarly as the thermal expansion coefficient, the isobaric heat capacity is not independent. Once the isothermal compressibility, the thermal pressure coefficient, and the isochoric heat capacity are known, it can be found from the Meyer relation:
\begin{equation}\label{eq:9}
	{C_P} = TV {\kappa_T \beta_V^2} + C_V,
\end{equation}
reflecting the combined thermal and mechanical response of the system.
By analogy to the isochoric heat capacity, we introduce the dimensionless isobaric heat capacity per particle by
\begin{equation}\label{eq:6}
	C^*_P = \frac{C_P}{k_{\rm B} \langle N \rangle}
\end{equation}
and the dimensionless isobaric heat capacity per cell by
\begin{equation}\label{eq:7}
	c^*_P = \frac{C_P}{k_{\rm B} N_v}.
\end{equation}
The connection between $c^*_P$ and $C^*_P$ is 
\begin{equation}\label{eq:8}
	c^*_P = \rho^* C^*_P.
\end{equation}
Given that the total volume of the system is $V = v N_v$, the final equation for $c_P^*$ in terms of the dimensionless quantities reads
\begin{equation}\label{eq:cp}
	{c^*_P} = T^* {\kappa^*_T \beta^{*2}_V} + c^*_V.
\end{equation}
\begin{figure}[h]
	\centering
	\includegraphics[width=0.465\textwidth]{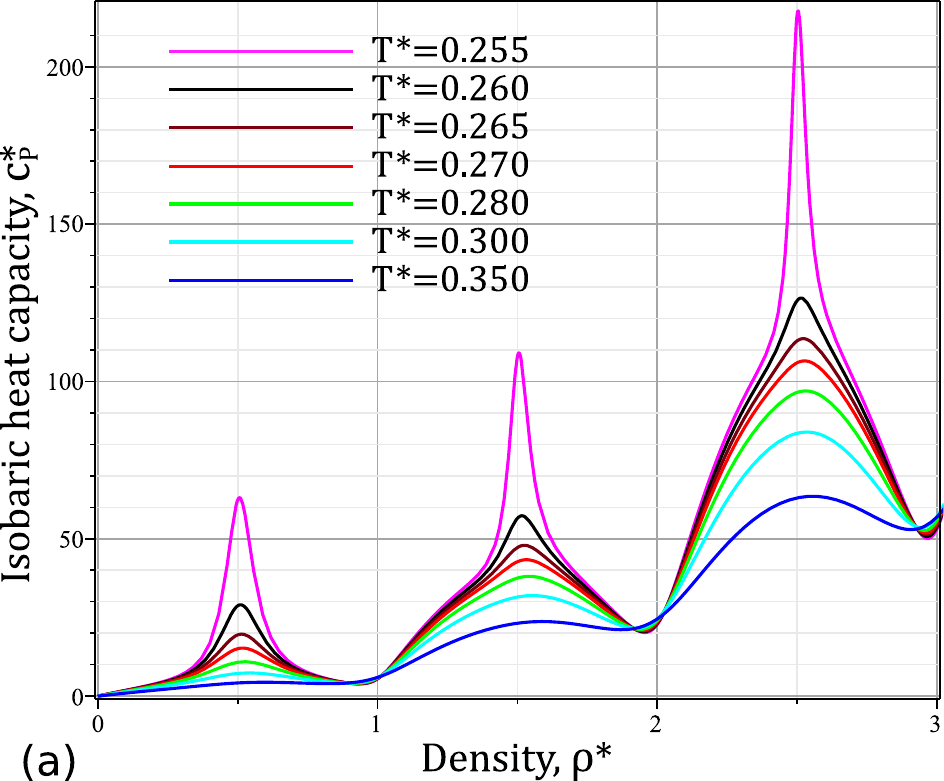}
	\includegraphics[width=0.465\textwidth]{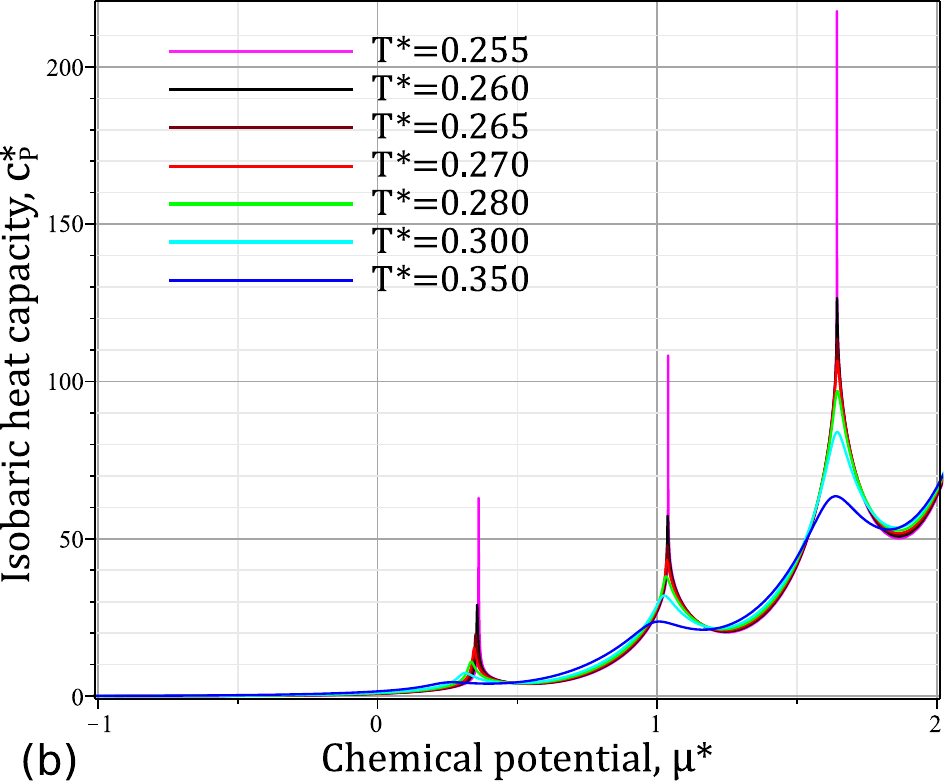}
	\captionsetup{width=0.9\textwidth}
	\caption{The isobaric heat capacity $c^*_P$ as a function of the particle number density $\rho^*$ (a) and the chemical potential $\mu^*$ (b) at different values of temperature $T^*$ in the supercritical region. Parameters taken as $f=1.5$, $v^*= 5.0$. }
	\label{fig7}
\end{figure}
Figure~\ref{fig7} shows the dependence of the isobaric heat capacity per cell $c^*_P$ on the particle number density $\rho^*$ and the chemical potential $\mu^*$ for several different values of supercritical temperature $T^*$. 

The calculated isobaric heat capacity remains strictly positive throughout the thermodynamically stable region. This behavior is consistent with the thermodynamic stability of the equilibrium states and indicates that heating the system at constant pressure always increases its enthalpy, namely, requires a positive heat input.

In contrast to the isochoric heat capacity, the isobaric heat capacity diverges at each critical point. This critical behavior follows directly from the Meyer relation. Since the isothermal compressibility diverges while the thermal pressure coefficient remains finite, the correction term $T^* {\kappa^*_T \beta^{*2}_V}$ relating $c_P^*$ and $c_V^*$ becomes singular. The divergence of $c_P^*$ is induced by the divergence of the isothermal compressibility.

The pronounced maxima observed in Fig.~\ref{fig7} therefore have a clear thermodynamic origin. As the critical point is approached, the mechanical response of the system becomes increasingly strong, making it possible for an infinitesimal temperature change at constant pressure to induce a substantial density variation. Additional thermal energy is then required not only to increase the internal energy but also to accommodate the accompanying volume change, leading to the divergence of the isobaric heat capacity.

The qualitative difference between the two heat capacities reflects the different thermodynamic constraints under which the system is heated. At constant volume, the density is fixed and the energetic response remains finite even at criticality. At constant pressure, however, the density is free to adjust to the external conditions, allowing the mechanical response encoded in the isothermal compressibility to manifest itself through the divergence of $c_P^*$.

\begin{figure}[h]
	\centering
	\includegraphics[width=0.465\textwidth]{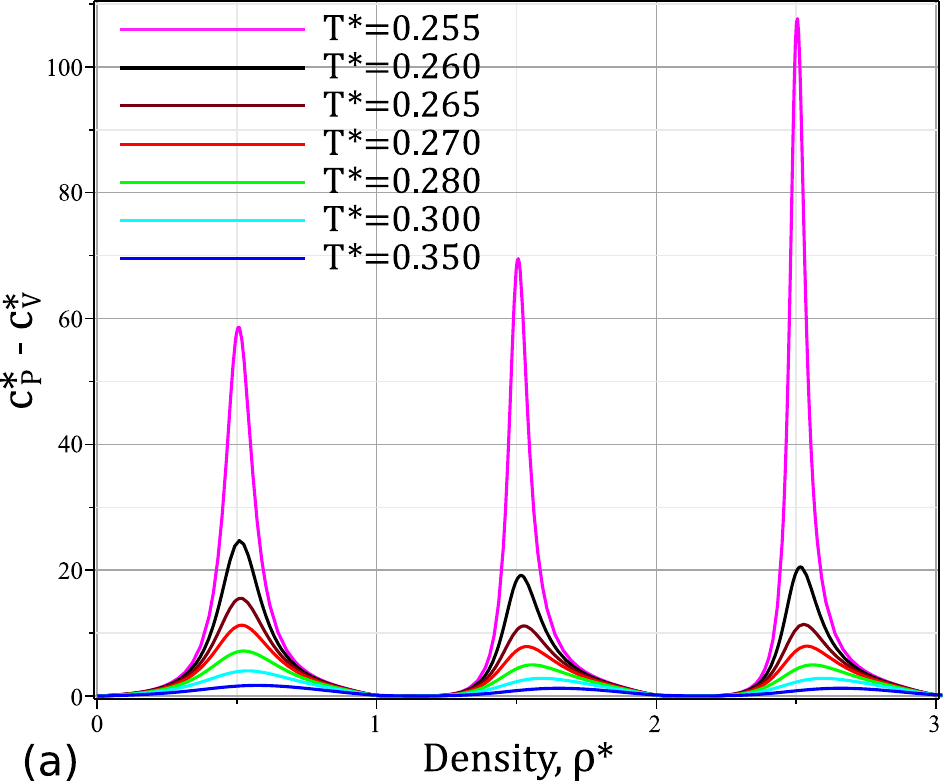}
	\includegraphics[width=0.465\textwidth]{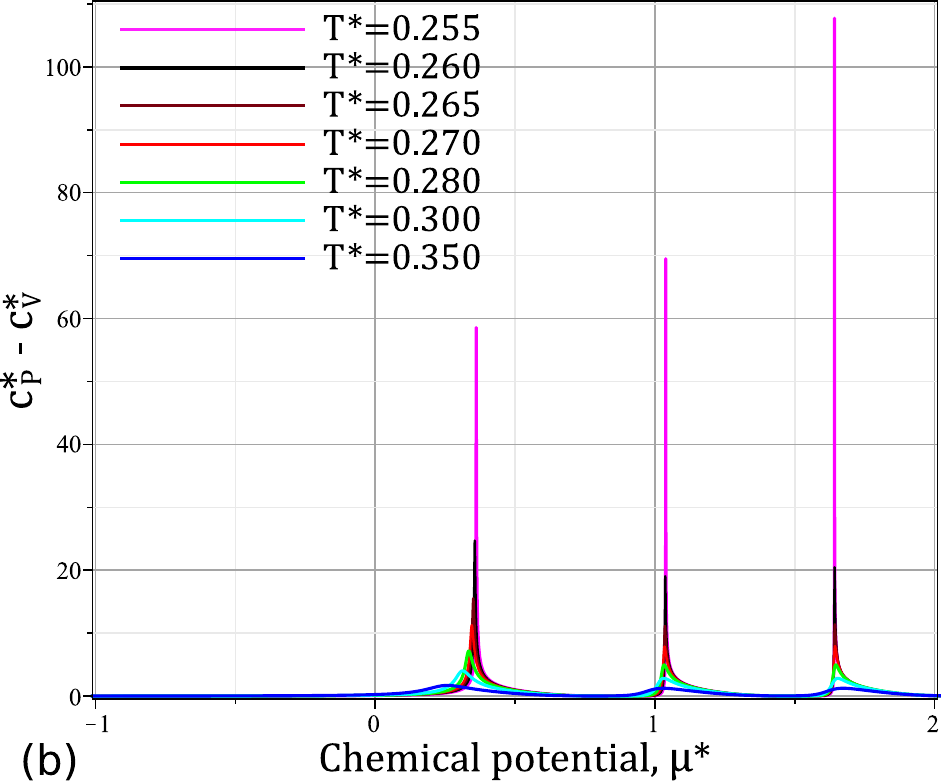}
	\captionsetup{width=0.9\textwidth}
	\caption{The difference between the isobaric and isochoric heat capacities $c^*_P - c^*_V$ as a function of the particle number density $\rho^*$ (a) and the chemical potential $\mu^*$ (b) at different values of temperature $T^*$ in the supercritical region. Parameters taken as $f=1.5$, $v^*= 5.0$.  }
	\label{fig8}
\end{figure}

The difference between the isobaric and isochoric heat capacities $c^*_P - c^*_V =T^* {\kappa^*_T \beta^{*2}_V} $ from \eqref{eq:cp} as a function the particle number density $\rho^*$ and the chemical potential $\mu^*$ at supercritical temperatures $T^*$ is displayed in Fig.~\ref{fig8}. Near each critical point, even a small temperature increase at constant pressure produces a substantial change in the equilibrium density owing to the divergent compressibility. 
Away from the critical points, the difference $c_P^*-c_V^*$ decreases rapidly and remains relatively small over most of the supercritical region. This behavior indicates that, sufficiently far from criticality, the thermal response of the system becomes nearly independent of the thermodynamic constraint under which heating is performed. In other words, the distinction between constant-pressure and constant-volume heating becomes progressively less significant as the mechanical susceptibility of the system decreases.


The behavior of the isobaric heat capacity obtained for the present CFM is qualitatively consistent with that reported for other model fluids. In the van der Waals fluid, $C_V$ is identically constant, so the isobaric heat capacity is given entirely by the Meyer correction, which diverges at a critical point, and the locus of $C_P$ maxima along isotherms collapses onto the single isochore~\cite{Johnston14,WAP24}. In the present model, by contrast, both $c_V^*$ and the correction term $T^*\kappa_T^*\beta_V^{*2}$ contribute non-trivially to $c_P^*$, and an infinite sequence of divergences appears, one at each critical density $\rho^{*(k)}_c$. 
In Lennard--Jones systems $C_P$ exhibits both maxima and minima in the supercritical region~\cite{YS13,MM14,MS13} and diverges at the critical point in accord with 3D Ising universality~\cite{WIH12,W95}, while the difference $C_P - C_V$ diverges at criticality and falls to near-ideal-gas values away from it~\cite{NOBAB17}. 
In purely repulsive soft-matter models, such as the Gaussian core model (GCM), $C_P$ and $C_P - C_V$ remain finite throughout the fluid range. The heat capacity at constant pressure increases sharply near the solid–liquid transition line and then decreases monotonically with increasing temperature exhibiting a pronounced minimum~\cite{LS19,MS11,MM12}. 
The double-Gaussian core model (DGM), which recovers a gas–liquid transition, displays both maxima and minima in $C_P$ above $T_c$~\cite{LS19}. 
For square-well fluids of variable range, $C_P$ diverges at the gas–liquid critical point and exhibit maxima in the supercritical region~\cite{SN11}. 
The distinctive feature of the Curie-Weiss cell fluid is therefore the analytically derived infinite sequence of $c_P^*$ divergences and $\left(c_P^* - c_V^*\right)$ peaks, whose locations are determined exactly by the critical densities $\rho^{*(k)}_c$ and whose thermodynamic origin is fully transparent through \eqref{eq:cp}.

%
%
%
%
%

\section{Conclusions}

In this work, we have presented a comprehensive analytical investigation of the thermodynamic response functions of the three-dimensional multiple-occupancy cell fluid model with Curie--Weiss interaction in the supercritical region. Explicit analytical expressions have been derived for the three independent thermodynamic response functions: the isothermal compressibility, the thermal pressure coefficient, and the isochoric heat capacity, from which the thermal expansion coefficient and the isobaric heat capacity follow directly through exact thermodynamic identities.  This treatment was possible owing to the knowledge of the analytical form of the equation of state, thereby eliminating the need for numerical differentiation. 

The response functions exhibit markedly different thermodynamic behavior. The isothermal compressibility remains positive throughout the stable region and diverges at every critical point, reflecting the loss of mechanical stability of the equilibrium state. The thermal pressure coefficient, in contrast, remains finite at criticality, exhibits a pronounced oscillatory dependence on density, and becomes negative within a limited interval of temperatures and densities, giving rise to negative values of the thermal expansion coefficient. The isochoric heat capacity also remains finite at the critical points despite developing pronounced maxima in their vicinity, whereas the isobaric heat capacity inherits the critical divergence of the isothermal compressibility through the Meyer relation. The difference between the two heat capacities provides a direct measure of the increasing influence of the mechanical response on the thermal properties as the critical points are approached.

Our results show that the different thermodynamic response functions capture distinct aspects of the equilibrium thermodynamic susceptibility of the model. While some quantities are governed primarily by the mechanical response of the system, others are determined by the interplay between thermal and interaction contributions to the equation of state.

The present study completes the analytical description of the thermodynamic response functions in the supercritical region and provides the theoretical foundation for a subsequent investigation in the subcritical region, where we will analyze phase coexistence, metastability, and the sequence of first-order fluid--fluid phase transitions predicted by the model.

\subsubsection*{Acknowledgement}

The authors are deeply grateful to all warriors of the Ukrainian Armed Forces, living and fallen, for making it possible to continue the research work.
The work was supported by the National Research Foundation of Ukraine, under the Project No.\ 2023.03/0201.

%
%
%
%
%
%

\appendix
\renewcommand{\theequation}{A.\arabic{equation}}
\setcounter{equation}{0}
\section{\label{sec:app-a}Phase diagram of the Curie--Weiss cell fluid model}

The cell fluid model exhibits a hierarchy of first-order phase transitions at sufficiently low temperatures \cite{KD22,DKPP26}.  

\begin{figure}[h!]
	\centering
	\includegraphics[width=0.49\textwidth]{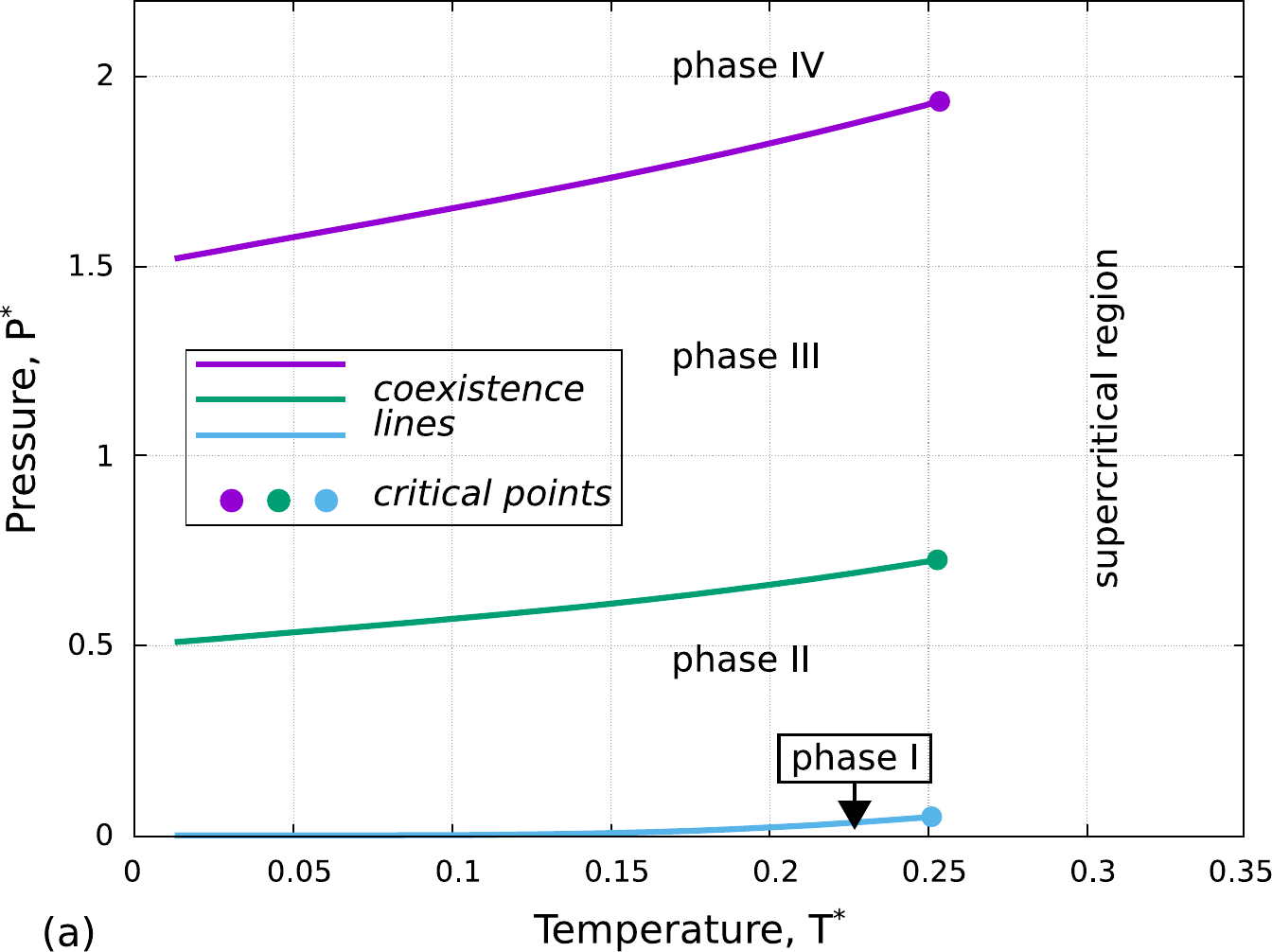} 
	\includegraphics[width=0.49\textwidth]{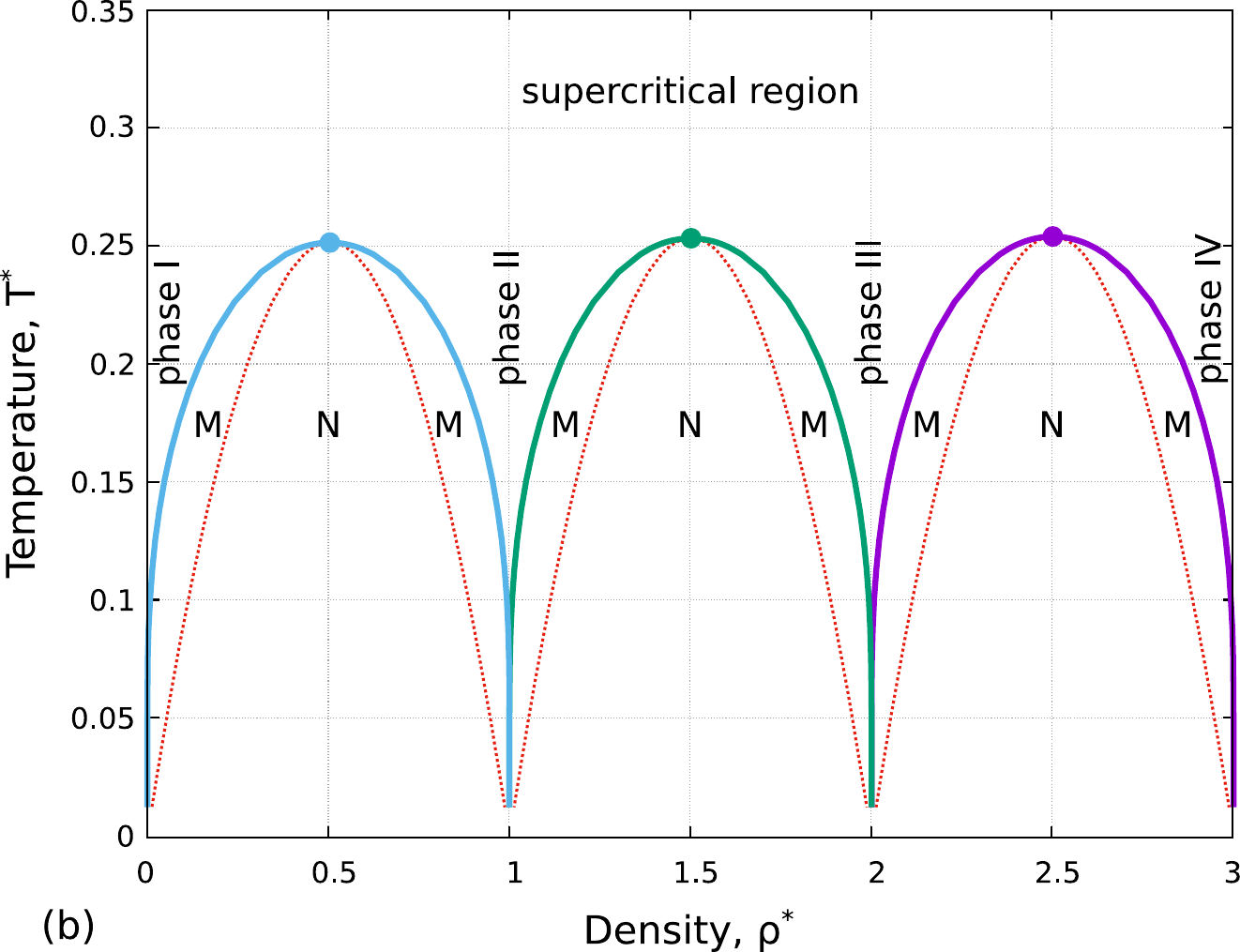}
	\caption{Phase diagrams in the temperature-pressure (a) and density-temperature (b) planes for the cell model with Curie-Weiss-type interaction potential, in case we set the repulsion to attraction ratio $f = 1.5$.  All diagrams are valid for arbitrary $v^*$. 
		\\ (a): ($P^*,T^*$) phase diagram showing the first three first-order phase transitions and the corresponding phase coexistence lines among four distinct phases. Each coexistence line terminates at its critical point (solid circles), where the supercritical region starts.  
		\\ (b): ($T^*,\rho^*$) phase diagram displaying spinodals (thin dotted red lines) and phase coexistence lines: the blue solid line indicates coexistence between Phase I and Phase II, the green solid line indicates coexistence between Phase II and Phase III, and the purple solid line indicates coexistence between Phase III and Phase IV. Regions marked by M and N denote metastable and unstable states, respectively.	
	}\label{fig0}
\end{figure}

Figure~\ref{fig0} presents the pressure-temperature and temperature-density phase diagrams of the cell fluid model with Curie--Weiss interaction. The model exhibits an infinite sequence of first-order phase transitions, each terminating at a corresponding critical point. For clarity, only the first three critical points are shown. The values of the temperature, the particle number density, the pressure and the chemical potential at the first three critical points are presented in Table~\ref{tab1}. The critical temperatures increase slightly with the critical point index $k$, while the critical densities are located approximately at $\rho^*\approx0.5$, $1.5$, and $2.5$. Above the coexistence curves, the system exists as a single supercritical phase.

\begin{table}[h!]
	\caption{Dimensionless parameters of the critical points: the critical temperature $T^*_c$, particle number density at criticality $\rho^*_c$, critical pressure $P^*_c$ and chemical potential $\mu^*_c$. The table includes data related to the first three critical points numbered by $k=1,2,3$. Numerical values are obtained for the ratio $f=1.5$; $\mu^{*(k)}_c = \mu^{*(k)}_c (v^*=5)$.  }
	\tabcolsep4.5pt
	\label{tab1}
	\begin{center}
		\begin{tabular}{|c|c|c|c|c|}
			\hline
			$k$	&   $T^{*(k)}_c$    &   $\rho^{*(k)}_c$     &  $P^{*(k)}_c$     &   $\mu^{*(k)}_c$    \\
			\hline
			1	&     0.251285  &   0.503869   &    0.0488583  &    0.365215  \\
			\hline
			2	&   0.253109    &    1.50136   &   0.725778  &     1.03938 \\
			\hline
			3	&   0.253838   &    2.50070   &   1.93280  &     1.64220 \\
			\hline
		\end{tabular}
	\end{center}
\end{table}

Since the present work is devoted exclusively to the supercritical region, the phase diagram serves as a reference for interpreting the behavior of the thermodynamic response functions discussed below. In particular, the locations of the critical points determine the density intervals where the response functions exhibit their most pronounced features. As it is shown in Sect.~\ref{sec:respons_functions}, the isothermal compressibility, thermal expansion coefficient, and isobaric heat capacity diverge at these critical points, whereas the thermal pressure coefficient and the isochoric heat capacity remain finite despite displaying significant changes in their behavior.

The repeated appearance of critical points is a distinctive feature of the present cell fluid model. Consequently, the response functions exhibit a corresponding sequence of characteristic anomalies, reflecting the underlying hierarchy of phase transitions and critical states. Throughout this work, the phase diagram shown in Fig.~\ref{fig0} provides the thermodynamic framework for interpreting these anomalies in the supercritical region.

\renewcommand{\theequation}{B.\arabic{equation}}
\setcounter{equation}{0}
\section{Explicit expressions for derivatives}
\subsection{\label{sec:app-b} Isothermal compressibility}

In this Appendix, we derive the explicit expression for the isothermal compressibility $\kappa^*_T$ from~\eqref{eq:kappa_star_m1}: 
\begin{equation}
	\kappa^*_T  =  \frac{1}{{\rho^*}^2} \left(\frac{\partial \rho^*}{\partial \mu^{*}}\right)_{T^*} .\nonumber
\end{equation}
Applying the chain rule to the parametric representation of the thermodynamic variables through $\bar{z}_{\rm max}$ yields
\begin{equation}\label{b1}
	\left(\frac{\partial \rho^*}{\partial \mu^{*}}\right)_{T^*} = \left(\frac{\partial \rho^*}{\partial \bar{z}_{\rm{max}}}\right)_{T^*} \left(\frac{\partial \mu^{*}}{\partial\bar{z}_{\rm{max}}}\right)_{T^*}^{-1} .
\end{equation}
For the particle number density $\rho^*$ from~\eqref{1d11} we have
\begin{equation}\label{b2}
	\left(\frac{\partial \rho^*}{\partial \bar{z}_{\rm{max}}}\right)_{T^*} = M_2(T^*;\bar{z}_{\rm max}),
\end{equation}
where the function $M_2(T^*;\bar{z}_{\rm max})$ is defined in~\eqref{eq:m2}. The derivative of the chemical potential with respect to $\bar{z}_{\rm max}$ follows directly from~\eqref{1d10} and is given by
\begin{equation}\label{b3}
	\left(\frac{\partial \mu^{*}}{\partial\bar{z}_{\rm{max}}}\right)_{T^*} = T^* - M_2(T^*;\bar{z}_{\rm max}).
\end{equation}
Substituting Eqs.~\eqref{b2} and~\eqref{b3} into Eq.~\eqref{b1} immediately yields the explicit expression for the isothermal compressibility given in Eq.~\eqref{eq:kappa_exp}.

\subsection{\label{sec:app-c} Thermal pressure coefficient}

This Appendix contains the explicit derivation of the thermal pressure coefficient $\beta^*_V$ from~\eqref{eq:beta_star_m}:

\begin{equation}
	\label{c1}
	\beta^*_V = \left(\frac{\partial P^*}{\partial T^*}\right)_{\mu^*} 
	- \left(\frac{\partial P^*}{\partial \mu^*}\right)_{T^*} 
	\left(\frac{\partial \rho^*}{\partial T^*}\right)_{\mu^*}
	\left(\frac{\partial \rho^*}{\partial \mu^*}\right)^{-1}_{T^*} . \nonumber
\end{equation}
The evaluation of Eq.~\eqref{eq:beta_star_m} requires four partial derivatives. We begin with the derivative of the pressure with respect to temperature at fixed volume and chemical potential which is actually the entropy per cell $S^*_v = \left(\dfrac{\partial P^*}{\partial T^*}\right)_{\mu^*} $ :
\begin{equation}\label{c2}
	\left(\frac{\partial P^*}{\partial T^*}\right)_{\mu^*} = E(T^*;\bar{z}_{\rm max}) + T^* \left(\frac{\partial E(T^*;\bar{z}_{\rm max})}{\partial T^*}\right)_{\mu^*} .
\end{equation}
The temperature derivative of the function $E(T^*;\bar{z}_{\rm max})$ can be written as
\begin{equation}\label{c3}
	 \left(\frac{\partial E(T^*;\bar{z}_{\rm max})}{\partial T^*}\right)_{\mu^*} =  \left(\frac{\partial }{\partial T^*} \left[ \ln K_0(T^*;\bar{z}_{\rm max})-\frac{1} {2T^*}\left[\frac{K_1(T^*;\bar{z}_{\rm max})}{K_0(T^*;\bar{z}_{\rm max})}\right]^2 \right] \right)_{\mu^*}.
\end{equation}
To evaluate this derivative one needs the derivatives of the functions $K_j$ with respect to temperature,
\begin{equation}\label{c4}
	 \left(\frac{\partial K_j(T^*;\bar{z}_{\rm max})}{\partial T^*}\right)_{\mu^*} = \frac{f}{2 T^{*2}} K_{j+2}(T^*;\bar{z}_{\rm max}) + K_{j+1}(T^*;\bar{z}_{\rm max}) \left(\frac{\partial \bar{z}_{\rm max}}{\partial T^*}\right)_{\mu^*}.
\end{equation}
where the derivative of the saddle-point parameter is
\begin{align}\label{c5}
	 \left(\frac{\partial \bar{z}_{\rm max}}{\partial T^*}\right)_{\mu^*} & = \left( T^* - M_2 (T^*;\bar{z}_{\rm max}) \right)^{-1} \left( \frac{3}{2}\left( \ln T^* + 1 \right) + \ln v^* - \bar{z}_{\rm max} \right . \nonumber \\
	& \left . + \frac{f}{2 T^{*2}} \left( \frac{K_3(T^*;\bar{z}_{\rm max})}{K_0(T^*;\bar{z}_{\rm max})} - M_1(T^*;\bar{z}_{\rm max}) \frac{K_2(T^*;\bar{z}_{\rm max})}{K_0(T^*;\bar{z}_{\rm max})} \right) \right).
\end{align}
Combining Eqs.~\eqref{c3}--\eqref{c5} leads to the explicit result
\begin{equation}\label{c6}
	\left(\frac{\partial P^*}{\partial T^*}\right)_{\mu^*} = \ln K_0(T^*;\bar{z}_{\rm max}) +\left(\frac{3}{2} - \bar{z}_{\rm max} \right) M_1(T^*;\bar{z}_{\rm max}) + \frac{f}{2 T^{*}} \frac{K_2(T^*;\bar{z}_{\rm max})}{K_0(T^*;\bar{z}_{\rm max})}.
\end{equation}
We next evaluate the derivative of the pressure with respect to chemical potential:
\begin{equation}\label{c7}
	\left(\frac{\partial P^*}{\partial \mu^*}\right)_{T^*} = T^* \left(\frac{\partial }{\partial \mu^*}  \left[ \ln K_0(T^*;\bar{z}_{\rm max})-\frac{1} {2T^*}\left[\frac{K_1(T^*;\bar{z}_{\rm max})}{K_0(T^*;\bar{z}_{\rm max})}\right]^2 \right]   \right)_{T^*}.
\end{equation}
For this purpose, the derivatives of the functions $K_j$ with respect to $\mu^*$ are required:
\begin{equation}\label{c8}
	\left(\frac{\partial K_j(T^*;\bar{z}_{\rm max})}{\partial \mu^*} \right)_{T^*} = K_{j+1}(T^*;\bar{z}_{\rm max}) \left(\frac{\partial \bar{z}_{\rm max}}{\partial \mu^*}\right)_{T^*}.
\end{equation}
Using the saddle-point relation one obtains
\begin{equation}\label{c9}
	 \left(\frac{\partial \bar{z}_{\rm max}}{\partial \mu^*}\right)_{T^*} = \frac{1}{ T^* - M_2(T^*;\bar{z}_{\rm max})}
\end{equation}
and therefore
\begin{equation}\label{c10}
		\left(\frac{\partial K_j(T^*;\bar{z}_{\rm max})}{\partial \mu^*} \right)_{T^*} = \frac{ K_{j+1}(T^*;\bar{z}_{\rm max})}{ T^* - M_2(T^*;\bar{z}_{\rm max})}.
\end{equation}
Substitution into Eq.~\eqref{c7} yields the remarkably simple expression
\begin{equation}\label{c11}
	\left(\frac{\partial P^*}{\partial \mu^*}\right)_{T^*} =  M_1(T^*;\bar{z}_{\rm max}).
\end{equation}
The remaining derivatives entering Eq.~\eqref{eq:beta_star_m} are obtained from the density representation. Differentiation with respect to temperature gives
\begin{align}\label{c12}
		\left(\frac{\partial \rho^*}{\partial T^*}\right)_{\mu^*} & = \frac{f}{2 T^{*2}} \left( \frac{K_3(T^*;\bar{z}_{\rm max})}{K_0(T^*;\bar{z}_{\rm max})} - M_1(T^*;\bar{z}_{\rm max}) \frac{K_2(T^*;\bar{z}_{\rm max})}{K_0(T^*;\bar{z}_{\rm max})} \right) \nonumber \\ 
		& +  M_2(T^*;\bar{z}_{\rm max}) \left(\frac{\partial \bar{z}_{\rm max}}{\partial T^*}\right)_{\mu^*} .
\end{align}
while differentiation with respect to chemical potential results in
\begin{equation}\label{c13}
		\left(\frac{\partial \rho^*}{\partial \mu^*}\right)_{T^*} =  \frac{M_2(T^*;\bar{z}_{\rm max})}{ T^* - M_2(T^*;\bar{z}_{\rm max}) }.
\end{equation}
Substituting Eqs.~\eqref{c6}, \eqref{c11}, \eqref{c12}, and \eqref{c13} into Eq.~\eqref{eq:beta_star_m} yields the explicit expression for the thermal pressure coefficient.

\subsection{\label{sec:app-d} Isochoric heat capacity}

Here we present the intermediate steps leading to the explicit derivation of the isochoric heat capacity per cell $c^*_V$ from~\eqref{eq:cv_cell}:
\begin{equation}\label{d1}
		c^*_V = T^* \left( \left(\frac{\partial S^*_v}{\partial T^*}\right)_{\mu^*} - \left(\frac{\partial \mu^*}{\partial \rho^*}\right)_{T^*}\left(\frac{\partial \rho^*}{\partial T^*}\right)_{\mu^*} \left(\frac{\partial S^*_v}{\partial \mu^*}\right)_{T^*} \right). \nonumber
\end{equation}
The calculation therefore requires the derivatives of the entropy per cell $S_v^*$ with respect to temperature and chemical potential, as well as the derivative of the density with respect to temperature. Differentiating the entropy per cell with respect to temperature at fixed volume and chemical potential gives
\begin{align}\label{d2}
	\left(\frac{\partial S^*_v}{\partial T^*}\right)_{\mu^*} & = \frac{f^2}{4 T^{*3}} \left( \frac{K_4(T^*;\bar{z}_{\rm max})}{K_0(T^*;\bar{z}_{\rm max})} -  \left[ \frac{K_2(T^*;\bar{z}_{\rm max})}{K_0(T^*;\bar{z}_{\rm max})} \right]^2 \right) \nonumber \\
	& + \frac{f}{2 T^{*2}} \left(\frac{3}{2} - \bar{z}_{\rm max} \right) \left( \frac{K_3(T^*;\bar{z}_{\rm max})}{K_0(T^*;\bar{z}_{\rm max})} - M_1(T^*;\bar{z}_{\rm max}) \frac{K_2(T^*;\bar{z}_{\rm max})}{K_0(T^*;\bar{z}_{\rm max})} \right) \nonumber \\
	& + \left[ \frac{f}{2 T^{*}} \left( \frac{K_3(T^*;\bar{z}_{\rm max})}{K_0(T^*;\bar{z}_{\rm max})} - M_1(T^*;\bar{z}_{\rm max}) \frac{K_2(T^*;\bar{z}_{\rm max})}{K_0(T^*;\bar{z}_{\rm max})} \right) \right . \nonumber \\ 
	& \left . +  M_2(T^*;\bar{z}_{\rm max})\left(\frac{3}{2} - \bar{z}_{\rm max} \right)  \right] \left(\frac{\partial \bar{z}_{\rm max}}{\partial T^*}\right)_{\mu^*} .
\end{align}
The derivative of the entropy per cell with respect to chemical potential takes the form
\begin{align}\label{d3}
   \left(\frac{\partial S^*_v}{\partial \mu^*}\right)_{T^*} & = \frac{1}{ T^* - M_2(T^*;\bar{z}_{\rm max}) }  \left[ \frac{f}{2 T^{*}} \left( \frac{K_3(T^*;\bar{z}_{\rm max})}{K_0(T^*;\bar{z}_{\rm max})} - M_1(T^*;\bar{z}_{\rm max}) \frac{K_2(T^*;\bar{z}_{\rm max})}{K_0(T^*;\bar{z}_{\rm max})} \right) \right . \nonumber \\ 
   & \left . +  M_2(T^*;\bar{z}_{\rm max})\left(\frac{3}{2} - \bar{z}_{\rm max} \right)  \right].
\end{align}
Finally, the temperature derivative of the density is
\begin{align}\label{d4}
	\left(\frac{\partial \rho^*}{\partial T^*}\right)_{\mu^*} & = \frac{f}{2 T^{*2}} \left( \frac{K_3(T^*;\bar{z}_{\rm max})}{K_0(T^*;\bar{z}_{\rm max})} - M_1(T^*;\bar{z}_{\rm max}) \frac{K_2(T^*;\bar{z}_{\rm max})}{K_0(T^*;\bar{z}_{\rm max})} \right)  \nonumber \\  & +  M_2(T^*;\bar{z}_{\rm max})\left(\frac{\partial \bar{z}_{\rm max}}{\partial T^*}\right)_{\mu^*}.
\end{align}
Together with the expression for $\left(\dfrac{\partial\mu^*}{\partial\rho^*}\right)_{T^*}$ obtained in Appendix~\ref{sec:app-c}, these relations provide all the quantities necessary for evaluating Eq.~\eqref{eq:cv_cell}. Substituting the above derivatives into Eq.~\eqref{eq:cv_cell} yields the explicit expression for the isochoric heat capacity presented by \eqref{eq:cv_cell_ex}.

\bibliographystyle{JHEPm}
\bibliography{bibs}

\end{document}